\documentclass[showpacs,preprintnumbers,amsmath,amssymb,nofootinbib]{revtex4}

\usepackage{graphicx}

\newcommand{\be}{\begin{equation}}
\newcommand{\ee}{\end{equation}}
\newcommand{\bea}{\begin{eqnarray}}
\newcommand{\eea}{\end{eqnarray}}

\newcommand{\F}{{\mathcal F}}
\newcommand{\Fo}{{\mathcal F}_0}

\newcommand{\onlygrqc}[1]{{#1}}

\begin{document}

\title{\bf Optimal filtering of the LISA data}

\author{Andrzej Kr\'olak}
\email{A.Krolak@impan.gov.pl}
\altaffiliation [currently at: ]{Institute of Mathematics,
Polish Academy of Sciences, Warsaw, Poland}
\affiliation{Jet Propulsion Laboratory, California Institute of
             Technology, Pasadena, CA 91109}

\author{Massimo Tinto}
\email{Massimo.Tinto@jpl.nasa.gov}
\altaffiliation [also at: ]{Space Radiation Laboratory, California
                            Institute of Technology, Pasadena, CA 91125}
\affiliation{Jet Propulsion Laboratory, California Institute of
             Technology, Pasadena, CA 91109}

\author{Michele Vallisneri}
\email{Michele.Vallisneri@jpl.nasa.gov}
\affiliation{Jet Propulsion Laboratory, California Institute of
             Technology, Pasadena, CA 91109}

\date{\today} 
\begin{abstract}
The LISA time-delay--interferometry responses to a
gravitational-wave signal are rewritten in a form that accounts for
the motion of the LISA constellation around the Sun; the responses are given in closed analytic forms valid for any frequency in the band accessible to LISA. We then present a complete procedure, based on the principle of maximum likelihood, to search for stellar-mass binary systems in the LISA data. We define the required optimal filters, the amplitude-maximized detection statistic (analogous to the $\mathcal{F}$ statistic used in pulsar searches with ground-based interferometers), and discuss the false-alarm and detection probabilities.
We test the procedure in numerical simulations of gravitational-wave detection.
\end{abstract}

\maketitle

\section{Introduction}

The Laser Interferometer Space Antenna (LISA) is a deep-space mission
aimed at detecting and studying gravitational radiation in the millihertz
frequency band. A joint American and European project, it is expected to
be launched in the year 2018, and to start collecting scientific data
approximately a year later, after reaching its orbital
configuration of operation \cite{Folkner97}. LISA consists of three widely separated spacecraft, flying in a triangular, almost-equilateral configuration, and exchanging coherent laser beams.
In contrast to ground-based, equal-arm, gravitational-wave (GW) interferometers, LISA will have multiple readouts, corresponding to the six laser Doppler shifts measured between spacecraft. Modeling each spacecraft as carrying lasers, beam splitters, photodetectors, and drag-free proof masses on each of two optical benches, Armstrong, Estabrook, and Tinto \cite{AET99,ETA00,TEA02} showed that it is possible to combine, with suitable time delays, the six time series of the inter-spacecraft Doppler shifts and the six time series of the intra-spacecraft Doppler shifts (measured between adjacent optical benches) to cancel the otherwise overwhelming frequency fluctuations of the lasers ($\Delta \nu/\nu \simeq 10^{-13}/ \sqrt{\rm Hz}$), and the noise due to the mechanical vibrations of the optical benches (which could be as large as $\Delta \nu/\nu \simeq 10^{-16}/ \sqrt{\rm Hz}$). The strain sensitivity level that then becomes achievable, $h \simeq 10^{-21} / \sqrt{\rm Hz}$, is set by the buffeting of the drag-free proof masses inside each optical bench, and by the shot noise at the photodetectors. Several such laser-noise--free interferometric combinations are possible, and they show different couplings to gravitational waves and to the remaining system noises \cite{AET99,ETA00,TAE01,TEA02}.
The technique used to synthesize these combinations is known as time-delay i9nterferometry (TDI); in the case of a stationary array, it was shown that the space of all the possible TDI observables can be constructed by combining four generators \cite{AET99,DNV}.

Recently, it was pointed out \cite{S,CH,STEA,TEA03}
that the rotational motion of the LISA array around the Sun and the time dependence of light travel times introduced by the relative (shearing) motion of the spacecraft have the effect of preventing the suppression of laser frequency fluctuations, at least under the current stability requirements, to the level of the secondary noises \emph{in the TDI observables as derived for a stationary array}. This problem was addressed by devising new combinations that are capable of suppressing the laser frequency fluctuations below the secondary noises for a rotating LISA array \cite{S,CH}, and for a rotating and shearing LISA array \cite{STEA,TEA03}. In this context, the original stationary-array combinations are sometimes known as ``TDI 1.0'' (or \emph{first-generation TDI}), the rotating-LISA combinations as ``TDI 1.5,'' and the rotating and shearing-LISA combinations as ``TDI 2.0;'' following Ref.\ \cite{TEA03}, we refer to the last as \emph{second-generation} TDI.
Second-generation combinations are essentially finite differences of first-generation combinations, and as such they appear more complicated. However, they retain the same sensitivity to incoming GWs: this is because the corrections introduced in the original combinations by the changing array geometry are obviously important for laser frequency fluctuations, but they are negligibly small for the GW responses and for the secondary noises; thus, once laser frequency noise is removed, the second-generation observables become finite differences of the corresponding first-generation observables. At a fixed frequency, the ratio of GW response to secondary noises (and hence the sensitivity) is then unchanged.

The GW responses of the TDI combinations depend on the relative orientation of the LISA array with respect to the direction of propagation of the GW signal, on the strength and polarization of the signal, and on its frequency components. Analytic expressions for the TDI responses were first derived by Armstrong, Estabrook and Tinto \cite{AET99}, for a stationary LISA array.
A realistic model of LISA must however include the motion of the array around the Sun, which introduces slow modulations in the phase and amplitude of the GW responses (in addition, of course, to the modifications introduced by adopting second-generation TDI). For instance, the LISA responses to the sinusoidal signal emitted by a binary system are not simple sinusoids, but rather superpositions of many sinusoids of smaller amplitude.
To maximize the likelihood of source detection, these effects must be modeled in GW search algorithms, either by including the modulations in the theoretical models of the signals (i.e., the \emph{templates}), or by demodulating the LISA data for a given set of sky positions as the first step of data analysis \cite{CLdemod,Hdemod}.

In this paper we derive the response of the second-generation TDI observables to the GW signals generated by a binary system, and we describe how signal templates based on these responses can be used in a maximum-likelihood, matched-filtering framework to search for binaries and to estimate their parameters once they are found.
Other methods to analyze the LISA data for signals from binaries, implemented in the long-wavelength approximation, have been proposed in Refs.\ \cite{CL03,C03}.
We work in the solar-system--baricentric frame, and we follow closely the derivation given by Jaranowski, Kr\'olak, and Schutz \cite{JKS98} for continuous sources and ground-based detectors.
A similar formalism was used by Giampieri \cite{G97} to obtain the antenna pattern of an arbitrary orbiting interferometer, in the long-wavelength approximation. The response of an orbiting equal-arm--Michelson interferometer to a sinusoidal signal was worked out by Cutler \cite{C98}, again in the long-wavelength limit.
Seto \cite{seto} extended Cutler's formalism to high frequencies (and to noise-canceling observables), in the context of studying optimal-filtering parameter estimation for supermassive--black-hole binaries. Cornish, Rubbo, and Poujade \cite{CR03,RCP03} obtained general expressions valid in the entire LISA frequency band, and for arbitrary GW signals; these expressions are given as integrals over the LISA arms, and they provide the basic building blocks to assemble the TDI observables. 
By contrast, in this paper we work out explicit time-domain expressions for the LISA response to moderately chirping binary systems, for all the second-generation TDI combinations. These expressions are valid over the entire LISA frequency band, and they are written as linear combinations of four time-dependent functions; this linear structure facilitates the computation of matched filters and the design of optimal filtering algorithms.

This paper is organized as follows. In Sec.\ II we give a brief
overview of the derivation of the TDI responses to GWs for a stationary array, and we argue that the corrections introduced by 
the motion of the LISA array and by the time dependence of light travel times are negligibly small. Working in the solar-system--baricentric frame, we obtain general expressions for the GW responses of the Michelson ($X_1$, $X_2$, $X_3$), Sagnac ($\alpha_1$, $\alpha_2$, $\alpha_3$), and optimal ($\bar{A}$, $\bar{E}$, $\bar{T}$; see \cite{PTLA02}) second-generation TDI observables \onlygrqc{(expressions for the first-generation observables are given in Appendix \ref{app:f})}; finally, we derive the corresponding closed-form analytic expressions for moderately chirping binary systems, valid at any frequency in the LISA band. In Sec.\ III we provide expressions for the spectral densities of noise in the TDI combinations. In Sec.\ IV we combine the results of Secs.\ II and III to design optimal filters that can be applied to the LISA TDI data to search for binary stars; we take advantage of the linear structure of the responses to define an optimal detection statistic that does not depend on the effective polarization and on the initial phase of the binary, in analogy to the $\mathcal{F}$ statistic \cite{JKS98,JK00}
used in searches for continuous GW sources with ground-based interferometers. In Sec.\ V we derive the false-alarm and false-dismissal probabilities for our LISA $\mathcal{F}$ statistic. Last, in Sec.\ VI we describe an efficient algorithm to compute $\mathcal{F}$, and we implement it numerically; we perform a simulation of GW detections in both the low and high-frequency part of the LISA band, and for both Michelson and optimal \cite{PTLA02} TDI observables, and we show that our algorithm yields very accurate estimates of source parameters. In the rest of this paper we shall use units where $c = 1$.

\section{Time-Delay Interferometry}

\begin{figure}
\centering
\includegraphics[angle=270,width=3.5in]{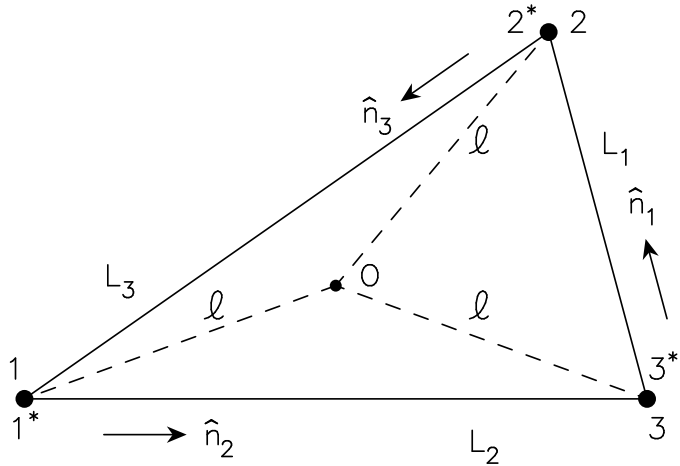}
\caption{Schematic LISA configuration.  The spacecraft are labeled 1, 2, and 3; each spacecraft contains two optical benches, denoted by 1, $1^*$, \ldots, as indicated.  The optical paths are denoted by $L_i$, where the index $i$ corresponds to the opposite spacecraft. The unit vectors $\hat{{\bf n}}_i$ point between pairs of spacecraft, with the orientation indicated.
\label{fig:geometry}}
\end{figure}

Figure \ref{fig:geometry} shows the overall LISA geometry.  The spacecraft are labeled 1, 2, and 3; the arms are labeled with the index of the opposite spacecraft (e.g., arm 1 lies between spacecraft 2 and 3). The light travel time (or, loosely, the \emph{armlength}) along arm $i$ is denoted by $L_i$ \cite{S,STEA,CH,TEA03}.
The basic constituents of the TDI observables are the time series of the relative laser-frequency fluctuations measured between spacecraft, which are denoted by $y_{ij}(t)$, with $i \ne j$: for instance,
$y_{31}(t)$ is the time series of relative frequency fluctuations measured for reception at spacecraft 1 with transmission from
spacecraft 3 (along arm 2); similarly, $y_{21}(t)$ is the
time series measured for reception at spacecraft 1
with transmission from spacecraft 2 (along arm 3), and so on.
Six more time series result from comparing the laser beams exchanged between adjacent optical benches within each spacecraft; these time series are 
denoted by $z_{ij}$, with $i,j = 1, 2, 3, \; i \ne j$ (see \cite{ETA00,TEA02,TEA03} for details). Delayed time series are denoted by commas: for instance, $y_{31,2} = y_{31}[t - L_2(t)]$, and so on.

The frequency fluctuations introduced by the lasers, by the optical
benches, by the proof masses, by the fiber optics, and by the measurement
itself at the photo-detector (i.e., the shot-noise fluctuations) enter the Doppler observables $y_{ij}$ and $z_{ij}$ with specific time signatures; see Refs.\ \cite{ETA00,TEA02,TEA03} for a detailed discussion.
The contribution $y^\mathrm{GW}_{ij}$ due to GW signals was derived in Ref.\ \cite{AET99} in the case of a stationary array. (Note that in Ref.\ \cite{AET99}, and indeed in all the literature on first-generation TDI, the notation $y_{ij}$ indicates the one-way Doppler measurement for the laser beam received at spacecraft $j$ and \emph{traveling along arm} $i$. In this paper we conform to the notation used in Refs.\ \cite{S,CH,STEA,TEA03}).

Since the motion of the LISA array around the Sun introduces a difference between (and a time dependence in) the corotating and counterrotating light travel times, the exact expressions for the GW contributions to the various first-generation TDI
combinations will in principle differ from the expressions valid for a stationary array \cite{AET99}.  However, the magnitude of the corrections introduced by the motion of the array are proportional to the product between the time derivative of the GW amplitude and the difference between the actual light travel times and those valid for a stationary array.  At 1 Hz, for instance, the larger correction to the signal (due to the difference between the corotating and counterrotating light travel times) is two orders of magnitude smaller than the main signal. Since the amplitude of this correction scales linearly with the Fourier frequency, we can completely disregard this effect (and the weaker effect due to the time dependence of the light travel times) over the entire LISA band \cite{TEA03}.  Furthermore, since along the LISA orbit the three armlengths will differ at most by $\sim$ 1\%--2\%, the degradation in signal-to-noise ratio introduced by
adopting signal templates that neglect the inequality of the armlengths will be at most a few percent.  For these reasons, in what follows we shall derive the GW responses of various second-generation TDI observables by disregarding the differences in the delay times experienced by light propagating clockwise and counterclockwise, and by assuming the three LISA armlengths to be constant and equal to $L = 5 \times 10^6 \, \mbox{km} \simeq 16.67 \, \mbox{s}$ \cite{PPA98}.
These approximations, together with the treatment of the moving-LISA GW response discussed at the end of Sec.\ \ref{sec:gwresponse}, are essentially equivalent to the \emph{rigid adiabatic approximation} of Ref.\ \cite{RCP03}, and to the formalism of Ref.\ \cite{seto}.

\subsection{Geometry of the orbiting LISA array}

\begin{figure}
\centering
\includegraphics[width=6in, angle=0]{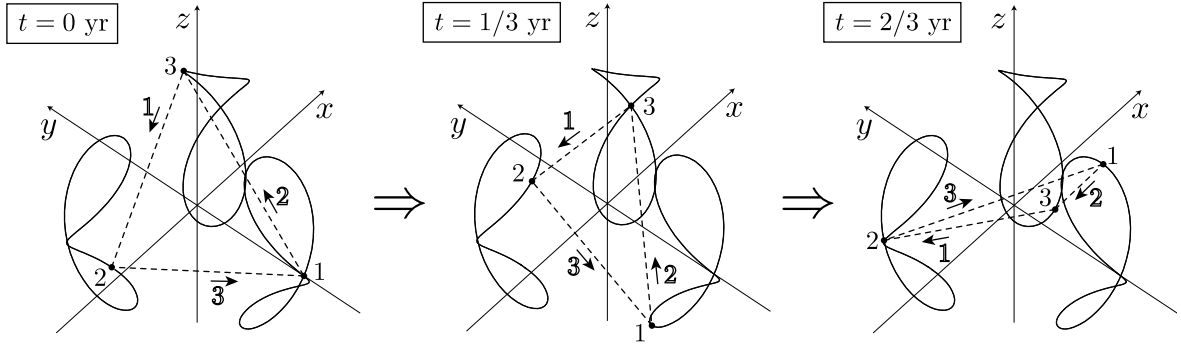}
\caption{Cartwheeling motion of the LISA array, as plotted in a frame with center in the LISA guiding center and axes parallel to the SSB ecliptic frame. We show three snapshots at different times along the LISA orbital period, 1 yr.\label{fig:lisacart}}
\end{figure}

We denote the positions of the three spacecrafts by $\mathbf{p}_i$ and the unit vectors along the arms by $\hat{\mathbf{n}}_i$, where $\hat{\mathbf{n}}_1$ points from spacecraft 3 to 2,
$\hat{\mathbf{n}}_2$ points from spacecraft 1 to 3, and 
$\hat{\mathbf{n}}_3$ points from spacecraft 2 to 1.
In the coordinate frame where the spacecraft are at rest, we can set without loss of generality
\begin{equation} 
{\bf p}^L_i = (L/\sqrt{3}) ( -\cos2\sigma_i , \sin2\sigma_i, 0),
\end{equation} 
and
\begin{equation} 
\hat{{\bf n}}^L_i = ( \cos\sigma_i , \sin\sigma_i, 0),
\label{eq:nlisa}
\end{equation} 
where
\begin{figure}
  \includegraphics[width=10cm]{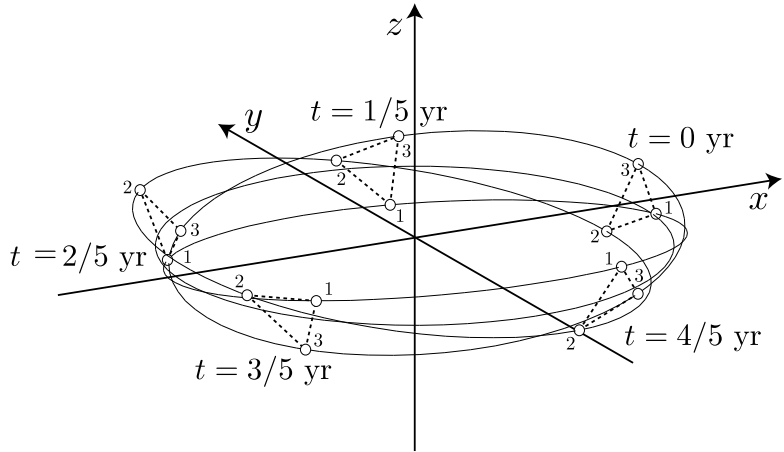}
  \caption{Orbital motion of the LISA detector, shown in a solar-system--baricentric ecliptic coordinate system. The trajectories shown correspond to setting $\zeta = -\pi/6$, $\Omega = 2\pi/$yr, and $\xi_0 = \eta_0 = 0$ in Eqs.\ \eqref{eq:guiding} and \eqref{eq:lisarot}.\label{fig:lisa}}
\end{figure}
\begin{equation} 
\label{eq:sigma}
\sigma_i = 3\pi/2 - 2(i-1)\pi/3.  
\end{equation}
Because the motion of the LISA guiding center (i.e., the baricenter of the formation) is contained in the plane of the ecliptic, it is convenient 
to work in a solar-system--baricentric (SSB) ecliptic coordinate system. We take the $x$ axis of this system to be directed toward the vernal point.
A realistic set of orbits for the spacecraft \cite{PPA98}, shown in Fig.\ \ref{fig:lisa}, is obtained by setting
\begin{equation}
{\bf p}_i(t) = {\bf r}(t) + {\sf O}_2 \cdot {\bf p}^L_i, \quad \hat{\bf n}_i(t) = {\sf O}_2 \cdot \hat{\bf n}^L_i,
\label{eq:poslisa}
\end{equation}
where {\bf r} is the vector from the origin of the SSB coordinate system to the LISA guiding center, as described by the SSB components
\begin{equation}
\label{eq:guiding}
{\bf r} = R (\cos\eta, \sin\eta, 0), \quad R = 1 \, \mathrm{AU};
\end{equation}
the function $\eta = \Omega t + \eta_0$ returns the true anomaly of the motion of the LISA guiding center around the Sun.
The rotation matrix ${\sf O}_2$ models the cartwheeling motions of the spacecraft along their inclined orbits, shown in Fig.\ \ref{fig:lisacart}; it is given by
\begin{equation}
{\sf O}_2 = \left(
\begin{array}{ccc}
 \sin\eta\cos\xi -\cos\eta\sin\zeta\sin\xi & -\sin\eta\sin\xi -\cos\eta\sin\zeta\cos\xi  & -\cos\eta\cos\zeta\\
-\cos\eta\cos\xi - \sin\eta\sin\zeta\sin\xi  & \cos\eta\sin\xi - \sin\eta\sin\zeta\cos\xi & -\sin\eta\cos\zeta\\
 \cos\zeta\sin\xi   &   \cos\zeta\cos\xi  &  -\sin\zeta\\
\end{array}
\right);
\label{eq:lisarot}
\end{equation}
the function $\xi = -\Omega t + \xi_0$ returns the phase of the motion of each spacecraft around the guiding center, while $\zeta$ sets the inclination of the orbital plane with respect to the ecliptic. For the LISA
trajectory, $\Omega = 2\pi/$yr and $\zeta = - \pi/6$ \cite{PPA98}. For simplicity, we can set $\eta_0 = \xi_0 = 0$, so that at time $t = 0$ the LISA guiding center lies on the positive $x$ axis of the SSB system, while ${\bf p}_1$ lies on the negative $y$ axis. The spacecraft orbits described by Eq.\ \eqref{eq:poslisa} can be approximately mapped to those used by Cornish and Rubbo \cite{CR03} by identifying our spacecraft 1, 2, and 3 with their spacecraft 0, 2, and 1, and by setting $\eta_0 = \kappa$, $\xi_0 = 3\pi/2 - \kappa + \lambda$, where $\kappa$ and $\lambda$ are the parameters defined below Eqs.\ (56) and (57) of Ref.\ \cite{CR03}.

\subsection{Generic plane waveform}

\begin{figure}
\includegraphics[width=4.5in]{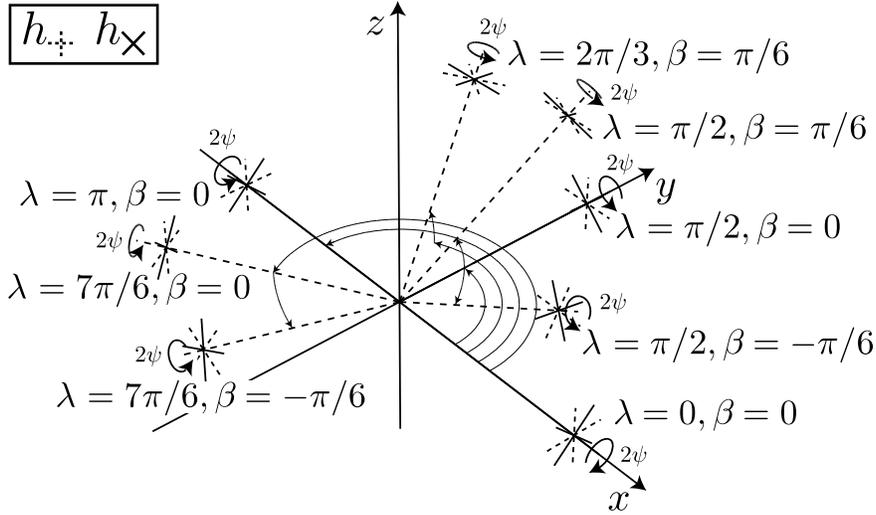}
\caption{Conventional definition of the GW polarizations $+$ (dashed) and $\times$ (solid) for various ecliptic latitudes $\beta$ and longitudes $\lambda$.\label{fig:polconv}}
\end{figure}

At the origin of the SSB frame, the transverse--traceless metric perturbation due to a source located at ecliptic latitude $\beta$ and longitude $\lambda$ can be written as
\begin{equation}
{\sf H}(t) = {\sf O}_1 \cdot {\sf H}^S(t) \cdot {\sf O}^{-1}_1,
\label{eq:hstr}
\end{equation}
where the metric perturbation in the source frame is taken to be
\begin{equation}
{\sf H}^S(t) = \left(%
\begin{array}{ccc}
  h^+(t) & h^\times(t) & 0 \\
  h^\times(t) & -h^+(t) & 0 \\
  0 & 0 & 0
\end{array}%
\right),
\label{eq:hst}
\end{equation}
with $h^+(t)$ and $h^\times(t)$ the two GW polarizations, and where
\begin{equation}
{\sf O}_1 = \left(
\begin{array}{ccc}
 \sin\lambda\cos\psi -\cos\lambda\sin\beta\sin\psi  & -\sin\lambda\sin\psi -\cos\lambda\sin\beta\cos\psi & -\cos\lambda\cos\beta\\
-\cos\lambda\cos\psi -\sin\lambda\sin\beta\sin\psi  &  \cos\lambda\sin\psi -\sin\lambda\sin\beta\cos\psi & -\sin\lambda\cos\beta\\
 \cos\beta\sin\psi                              &  \cos\beta\cos\psi                             & -\sin\beta\\
\end{array}
\right);
\end{equation}
the dependence of the rotation matrix ${\sf O}_1$ on $\beta$ and $\lambda$ enforces the transversality of the plane waves, which are propagating from a source located in the direction
\begin{equation}
\hat{{\bf k}} = (\cos\lambda\cos\beta, \sin\lambda\cos\beta, \sin\beta);
\end{equation}
the polarization angle $\psi$ encodes a rotation around the direction of wave propagation, $-\hat{{\bf k}}$, setting the convention used to define the two polarizations, $+$ and $\times$. The polarizations corresponding to $\psi = 0$ are shown in Fig.\ \ref{fig:polconv} for various source positions in the sky.
In the center of the LISA proper frame (the frame where the spacecraft are at rest), the transverse-traceless metric perturbation is given by
\begin{equation}
{\sf H}^L(t)  =  {\sf O}^{-1}_2(t) {\sf O}_1 {\sf H}^S(t) {\sf O}^{-1}_1 {\sf O}_2(t).
\label{HT}
\end{equation}
The time variable $t$ that appears in $h^+(t)$ and $h^\times(t)$ [and therefore in ${\sf H}(t)$ and ${\sf H^L}(t)$] is the time at the origin of the SSB frame. It is related to the time in the GW source frame by a relativistic time dilation, due to the proper motion of the source and to cosmological effects. It is however expedient to identify the two times, and to describe GW emission using SSB time; the time dilation is then taken into account by mapping the apparent (measured) physical parameters of a source into its real parameters.
The source positional parameters $\beta$, $\lambda$, and $\psi$ can be mapped to the parameters $\theta$, $\phi$, and $\psi$ used in Ref.\ \cite{CR03} by setting $\beta = \pi/2 - \theta$, $\lambda = \phi$, and $\psi = -\psi$.

\subsection{GW response of the LISA array}
\label{sec:gwresponse}

As derived in Ref.\ \cite{AET99} for a stationary, equilateral-triangle LISA array, the one-way Doppler responses $y_{21}$ and $y_{31}$ excited by a plane transverse--traceless GW propagating from the source direction $\hat{\mathbf{k}}$, are given by
\begin{eqnarray}
y_{21}^\mathrm{GW}(t) &=& [1 - \hat{\mathbf{k}} \cdot \hat{\mathbf{n}}_3][\Psi_3(t+\hat{\mathbf{k}} \cdot \mathbf{p}_2 - L)-\Psi_3(t+\hat{\mathbf{k}} \cdot \mathbf{p}_1)], \label{eq:tdiobsa} \\
y_{31}^\mathrm{GW}(t) &=& [1 + \hat{\mathbf{k}} \cdot \hat{\mathbf{n}}_2][\Psi_2(t+\hat{\mathbf{k}} \cdot \mathbf{p}_3 - L)-\Psi_2(t+\hat{\mathbf{k}} \cdot \mathbf{p}_1)]
\label{eq:tdiobs}
\end{eqnarray} 
[in the notation of Ref.\ \cite{AET99}, our $y_{21}$, $y_{31}$, and $\hat{\mathbf{k}}$ correspond to $y_{31}$, $y_{21}$, and $-\hat{\mathbf{k}}$ respectively] where 
\begin{equation}
\Psi_j(t) = \frac{\Phi_j(t)}{1 - (\hat{\mathbf{k}} \cdot \hat{\mathbf{n}}_j)^2}, \quad \Phi_j(t) = {\textstyle \frac{1}{2}} \, \hat{{\bf n}}'_j \cdot {\sf H}(t) \cdot \hat{{\bf n}}_j
\label{PHI_J}
\end{equation}
[the prime denotes vector transposition]. The two $\Psi_i$ terms in each of Eqs.\ \eqref{eq:tdiobsa} and \eqref{eq:tdiobs} correspond to the events of emission (at spacecraft 2 and 3, respectively) and reception (at spacecraft 1) of a laser photon packet; the time of the emission event is therefore retarded by an armlength $L$. The $\hat{\mathbf{k}} \cdot \mathbf{p}_i$ terms represent the retardation of the gravitational wavefronts to the positions of the spacecraft. The other four one-way Doppler responses are obtained by cyclical permutation of the indices ($1 \rightarrow 2$, $2 \rightarrow 3$, $3 \rightarrow 1$).

Our approximation to the GW response of the moving LISA array is obtained simply by interpreting Eqs.\ \eqref{eq:tdiobs} and \eqref{PHI_J} as written in the SSB ecliptic frame, and by adopting the time-dependent equations \eqref{eq:poslisa} for $\mathbf{p}_i$ and $\hat{\mathbf{n}}_i$.
Note that $\Phi_j(t)$ can then be written either as $\Phi_j(t) = {\textstyle \frac{1}{2}} \, \hat{\bf n}'_j(t) \cdot {\sf H}(t) \cdot \hat{\bf n}_j(t)$, or $\Phi_j(t) = {\textstyle \frac{1}{2}} \, (\hat{\bf n}_j^L)' \cdot {\sf H}^L(t) \cdot \hat{\bf n}_j^L$. The time-dependent rotation of the $\hat{\mathbf{n}}_i(t)$ introduces an amplitude modulation of the responses, generating sidebands at frequency multiples of $1/\mathrm{yr}$; the time dependence of the wavefront-retardation products $\hat{\mathbf{k}} \cdot \mathbf{p}_i(t)$ introduces a time-dependent Doppler shift caused by the relative motion of the spacecraft with respect to the SSB frame.

\subsection{Chirping-binary waveforms}

In the Newtonian limit, the GW signal emitted by a binary system 
located in the direction $\hat{\mathbf{k}}$ can be written in the form of Eq.\ \eqref{eq:hstr}, with 
\begin{equation}
h^+(t)      = h^+_0 \cos [\phi_s(t) + \phi_0], \quad
h^\times(t) = h^\times_0 \sin [\phi_s(t) + \phi_0].
\end{equation}
Here $\phi_0$ is an arbitrary constant phase, and the constant amplitudes $h^+_0$ and $h^\times_0$ are given by
\begin{equation}
h_0^+ = h_0 (1 + \cos^2 \iota) / 2,  
\quad
h_0^{\times} = h_0 \cos \iota,
\end{equation}
where $\iota$ is the angle between the normal to the orbital plane of the binary and the direction of propagation $-\hat{\mathbf{k}}$, and where
\begin{equation}
h_0 \simeq \frac{4 (G \mathcal{M}_c)^{5/3}}{c^4 D} \left[ \frac{\omega}{2} \right]^{2/3} \! \!, \end{equation}
with $\mathcal{M}_c = m_1^{3/5} m_2^{3/5} / (m_1 + m_2)^{1/5}$ the chirp mass, $\omega$ the angular frequency of the GW at $t=0$, and $D$ the luminosity distance to the source. Last, the phase $\phi_s(t)$ is given, to the first post--Newtonian order, by \cite{B02}
\begin{equation}
\phi_s (t) \simeq \tilde{\phi}_s(t) - \tilde{\phi}_s(0),
\quad
\tilde{\phi}_s (t) = 
 - \frac{2 M}{\mu}\Theta^{5/8}(t) \left[1 + \left(\frac{3715}{8064} + \frac{55}{96}\frac{\mu}{M} \right) \Theta^{-1/4}(t)\right],
\label{eq:phPN}
\end{equation}
where
\begin{equation}
\Theta(t) = \frac{\mu c^3}{5 G M^2} (t_c-t),
\end{equation}
\begin{equation}
t_c = \frac{G M^2}{\mu c^3}\frac{5}{256} \frac{1}{x_0^4} \left[1 + \left(\frac{743}{252} + \frac{924}{252}\frac{\mu}{M}\right)x_0\right], \quad
x_0 = \left[\frac{G M \omega}{2 c^3}\right]^{2/3},
\end{equation}
with $M = m_1 + m_2$ the total mass of the binary, and $\mu = m_1 m_2 / M$ the reduced mass. The time $t_c$ is the time to coalescence of the binary from the initial instant $t = 0$.
\begin{table}
\begin{center}
\begin{tabular}{c||c|c|c|c|c|c|c|c|c|c}
& \multicolumn{2}{c|}{$f = 10^{-3}$ Hz}         & \multicolumn{2}{c|}{$f =
2\times 10^{-2}$ Hz} &
                 \multicolumn{3}{c|}{$f = 5\times 10^{-2}$ Hz} &
\multicolumn{3}{c}{$f = 10^{-1}$ Hz} \\
Binary & N & 1PN  & N & 1PN   & N & 1PN & Doppler & N & 1PN & Doppler \\
\hline \hline
WD--WD  & 0 & 0 & $24^\dagger$         & 0    & \multicolumn{3}{c|}{-}
& \multicolumn{3}{c}{-}       \\ \hline
WD--NS  & 0 & 0 & $69^\dagger$         & 0    & \multicolumn{3}{c|}{-}
& \multicolumn{3}{c}{-}       \\ \hline
WD--BH  & 0 & 0 & $190^\dagger$        & 0    & \multicolumn{3}{c|}{-}
& \multicolumn{3}{c}{-}       \\ \hline
NS--NS  & 0 & 0 & $240^\dagger$        & 0    & $6.9\times 10^3$ & 3.4 & 0 &
$9.3\times 10^4$ & 78 & 2.7 \\ \hline
NS--BH  & 0 & 0 & $740$ & 0.33 & $2.2\times 10^4$ & 19.0 & 0.66 & $3.5\times
10^5$ & 640 & 8.5
\end{tabular}
\end{center}
\caption{Contributions to the evolution of GW frequency for various types of compact, stellar-mass binaries (white dwarfs with $m = 0.35 M_\odot$, neutron stars with $m = 1.4 M_\odot$, and black holes with $m = 6 M_\odot$), for selected (initial) GW frequencies within the LISA band. The contributions are expressed as GW cycles over one year of evolution, and the effects of Newtonian-order (N) and first post--Newtonian-order (1PN) terms are shown separately.
The column labeled ``Doppler'' reports the integrated phase shift (in cycles) due to the  increased Doppler shifting of the source as the frequency increases [see Eq.\ \eqref{eq:dopplershifting}], where significant. At $f = 10^{-3}$ Hz there is no significant evolution of GW frequency over one year. The symbol ``${}^\dagger$'' indicates that the Taylor expansion of the phase given by Eq.\ \eqref{eq:phT} is accurate to within a quarter of a cycle. Numbers are not shown where a binary of a given class cannot exist at a given frequency. Some of the conclusions that can be drawn from this table are apparent also in Figs.\ 10 and 12 of Ref.\ \cite{RCP03}: up to about 1 mHz, LISA cannot differentiate (using one year of data) between a monochromatic binary and a chirping binary (see Fig.\ 10 of Ref.\ \cite{RCP03}); above that frequency, chirping becomes appreciable (one additional GW cycle over a year in this table corresponds to a frequency shift of one bin in Fig.\ 12 of Ref.\ \cite{RCP03}), but we see that it can still be modeled faithfully by the linear-chirp model of Eq.\ \eqref{eq:phT}. 
\label{Tab:chirp}}
\end{table}

In Table \ref{Tab:chirp}, for binaries consisting of various combinations of white dwarfs (WDs, with $m = 0.35 M_\odot$), neutron stars (NSs, with $m = 1.4 M_\odot$), and black holes (BHs, with $m = 6 M_\odot$), and for various fiducial GW frequencies within the LISA band, we show the contributions to the evolution of GW frequency over one year caused by terms at the Newtonian (N) and first post--Newtonian (1PN) order. The table shows that at frequencies smaller or equal to $10^{-3}$ Hz, the evolution of frequency is negligible. At frequencies approaching 10 mHz, the change in frequency becomes significant, and needs to be included in the model of the signal; however, only the first derivative of the frequency is needed up to about 50 mHz. In binaries with WDs of mass $\sim 0.35 M_\odot$, above $\sim 20$ mHz the WDs fill their Roche lobe, and the dynamical evolution of the system is then determined by tidal interaction between the stars. In binaries with either a NS or a BH, post--Newtonian effects become important at about $\sim 50$ mHz. At $1$ Hz and above, these binaries will coalesce in less than $1$ yr; furthermore, population studies \cite{NYP00} suggest that the expected number of binaries above $50$ mHz containing neutron stars and black holes is negligible. (The effects of frequency evolution in the LISA response to GW signals from inspiraling binaries are also discussed in Ref.\ \cite{takahashiseto}.)

Therefore, for sufficiently small binary masses, for sufficiently small GW frequencies (and definitely for all non-tidally-interacting binaries that contain WDs), we can approximate the phase of the signal by Taylor-expanding it, and then neglecting terms of cubic and higher order. The resulting expression for the signal phase $\phi_s (t)$ is
\begin{equation}
\label{eq:phT}
\phi_s (t) \simeq \omega t + {\textstyle \frac{1}{2}}\dot{\omega} t^2,
\quad \mbox{where} \quad
\dot{\omega} = \frac{48}{5}\left(\frac{G {\cal M}_c}{2 c^3}\right)^{5/3}\!\!\!\omega^{11/3}.
\end{equation}

\subsection{TDI responses}
\label{sec:tdiresp}
The response of the second-generation TDI observables to a transverse--traceless, plane GW is obtained by setting $y_{ij}(t) = y^\mathrm{GW}_{ij}(t)$ [according to Eqs.\ \eqref{eq:tdiobsa} and \eqref{eq:tdiobs}] in the TDI expressions of Ref.\ \cite{STEA,TEA03}. For instance, the GW response of the second-generation TDI observable $X_1$ is given by
\begin{eqnarray}
X^\mathrm{GW}_1 & = &
\underbrace{ 
\left[
(y^\mathrm{GW}_{31} + y^\mathrm{GW}_{13,2}) 
+ (y^\mathrm{GW}_{21} + y^\mathrm{GW}_{12,3})_{,22} 
- (y^\mathrm{GW}_{21} + y^\mathrm{GW}_{12,3}) 
- (y^\mathrm{GW}_{31} +  y^\mathrm{GW}_{13,2})_{,33} 
\right]}_{X^\mathrm{GW}(t)}
\nonumber \\
& & 
- \underbrace{\left[
(y^\mathrm{GW}_{31} + y^\mathrm{GW}_{13,2}) 
+ (y^\mathrm{GW}_{21} + y^\mathrm{GW}_{12,3})_{,22} 
- (y^\mathrm{GW}_{21} + y^\mathrm{GW}_{12,3}) 
- (y^\mathrm{GW}_{31} +  y^\mathrm{GW}_{13,2})_{,33} 
\right]_{,2233}}_{X^\mathrm{GW}(t-2L_2-2L_3) \simeq X^\mathrm{GW}(t-4L)}.
\label{eq:30}
\end{eqnarray}
As anticipated above, here we are disregarding the effects introduced by the
time dependence of light travel times, and by the rotation-induced difference 
between clockwise and counterclockwise light travel times \cite{TDInote2}.
Each of the two terms delimited by square brackets in Eq.\ \eqref{eq:30} corresponds to the GW response of the first-generation Michelson observable $X$ \cite{AET99}. The TDI observables $X_2$ and $X_3$ are obtained by cyclical permutation of indices in Eq.\ \eqref{eq:30}. Likewise, the second-generation Sagnac observables $\alpha_1$, $\alpha_2$, and $\alpha_3$ can be written in terms of the first-generation Sagnac observables $\alpha$, $\beta$, and $\gamma$ \cite{STEA,TEA03}:
\begin{equation}
\alpha^\mathrm{GW}_1 (t) = \alpha^\mathrm{GW} (t) - \alpha^\mathrm{GW} (t - L_1 - L_2 - L_3) \simeq \alpha^\mathrm{GW} (t) - \alpha^\mathrm{GW} (t - 3 L).
\label{eq:32}
\end{equation}
We shall now assemble the Doppler measurements $y^\mathrm{GW}_{ij}$ from the various ingredients that enter Eqs.\ \eqref{eq:tdiobsa} and \eqref{eq:tdiobs}. We begin with the functions $\Phi_j$ of Eq.\ \eqref{PHI_J}, which can be rewritten as a linear combination of the two GW polarizations $h^+(t)$ and $h^\times(t)$:
\begin{equation}
\label{eq:defphi}
\Phi_j(t) = F_j^+(t) h^+(t)  + F_j^\times (t) h^\times (t),
\end{equation}
where
\begin{eqnarray}
F_j^+(t) &=& u_j(t)\cos 2\psi + v_j(t)\sin 2\psi, \\ 
F_j^\times(t) &=& v_j(t)\cos 2\psi - u_j(t)\sin 2\psi. 
\end{eqnarray}
The \emph{modulation functions} $u_i(t)$ and $v_i(t)$ depend rather intricately on the LISA-to-SSB (${\sf O}_2$) and source-to-SSB (${\sf O}_1$) rotations; thus, $u_i(t)$ and $v_i(t)$ depend on time through $\eta(t)$ and $\xi(t)$, and on the position of the source in the sky, given by the ecliptic coordinates $\beta$ and $\lambda$. Explicitly, we have
\begin{eqnarray}
u_i(t) &=& U_0\cos(-2 \gamma_i)
         + U_1\cos(\delta - 2 \gamma_i)
         + U_2\cos(2\delta - 2 \gamma_i)
         + U_3\cos(3\delta - 2 \gamma_i)
         + U_4\cos(4\delta - 2 \gamma_i) \label{ak} \\
       & & + \, ({\textstyle \frac{1}{4}} - {\textstyle \frac{3}{8}}\cos^2\zeta)\cos^2\beta
           - {\textstyle \frac{1}{2}}\sin\beta\cos\beta\cos\zeta\sin\zeta\cos\delta
           + {\textstyle \frac{1}{4}}\cos^2\zeta(1-{\textstyle \frac{1}{2}}\cos^2\beta)\cos 2\delta,          
          \nonumber \\[2mm]
v_i(t) &=& V_0\sin(-2 \gamma_i)
         + V_1\sin(\delta - 2 \gamma_i)
         + V_3\sin(3\delta - 2 \gamma_i)
         + V_4\sin(4\delta - 2 \gamma_i) \label{bk} \\
       & & - \, {\textstyle \frac{1}{2}}\cos\zeta\sin\zeta\cos\beta\sin \delta 
           + {\textstyle \frac{1}{4}} \cos^2\zeta\sin\beta\sin 2\delta 
           \nonumber,
\end{eqnarray}
where
\begin{eqnarray}
\label{eq:del}
\delta(t)   &=& \lambda - \eta(t) = \lambda - \eta_0 - \Omega t, \\
\label{eq:gai}
\gamma_i &=& \lambda - \eta_0 - \xi_0 - \sigma_i
\end{eqnarray}
[see Eq.\ \eqref{eq:sigma} for the definition of $\sigma_i$, and remember that $\eta = \Omega t + \eta_0$, $\xi = -\Omega t + \xi_0$], and where the coefficients $U_I$ and $V_I$ are given by
\begin{eqnarray}
U_0 &=& {\textstyle \frac{1}{16}} (1 + \sin^2\beta)(1 - \sin\zeta)^2,   \\
U_1 &=& -{\textstyle \frac{1}{8}} \sin 2\beta \cos\zeta(1 - \sin\zeta), \\
U_2 &=& {\textstyle \frac{3}{8}} \cos^2 \beta \cos^2\zeta,               \\
U_3 &=& {\textstyle \frac{1}{8}} \sin 2\beta \cos\zeta(1 + \sin\zeta),  \\
U_4 &=& {\textstyle \frac{1}{16}} (1 + \sin^2\beta)(1 + \sin\zeta)^2,   \\
\nonumber \\
V_0 &=& -{\textstyle \frac{1}{8}} \sin\beta(1 - \sin\zeta)^2,          \\
V_1 &=& {\textstyle \frac{1}{4}} \cos\beta\cos\zeta(1 - \sin\zeta),    \\
V_3 &=& {\textstyle \frac{1}{4}} \cos\beta\cos\zeta(1 + \sin\zeta),    \\
V_4 &=& {\textstyle \frac{1}{8}} \sin\beta(1 + \sin\zeta)^2,
\end{eqnarray}
with $\zeta = - \pi/6$. Expanding the antenna patterns $F^+_j(t)$ and $F^\times_j(t)$ of Eq.\ \eqref{eq:defphi}, and using trigonometric identities to absorb the initial phase $\phi_0$ into constant coefficients, the functions $\Phi_j(t)$ can be finally written as
\begin{equation}
\Phi_j(t) = a^{(1)} u_j(t) \cos\phi_s(t) + a^{(2)} v_j(t) \cos\phi_s(t) +
            a^{(3)} u_j(t) \sin\phi_s(t) + a^{(4)} v_j(t) \sin\phi_s(t),
\end{equation}
where the constant amplitudes $a^{(k)}$ are given by
\begin{eqnarray}
a^{(1)} &=&  \ \ h_0^+\cos\phi_0 \,\cos 2\psi -  h_0^\times\sin\phi_0 \,\sin 2\psi, \label{eq:ampone} \\
a^{(2)} &=&  \ \ h_0^+\cos\phi_0 \,\sin 2\psi +  h_0^\times\sin\phi_0 \,\cos 2\psi, \\
a^{(3)} &=& - \ h_0^+\sin\phi_0\,\cos 2\psi  -  h_0^\times\cos\phi_0\,\sin 2\psi, \\
a^{(4)} &=& - \ h_0^+\sin\phi_0\,\sin 2\psi  +  h_0^\times\cos\phi_0\,\cos 2\psi.
\label{Aamp}
\end{eqnarray}
Because the timescale of detector motion is much longer than the typical GW period (and because we are neglecting the evolution of the GW amplitude $h_0$), it is sufficient to apply the retardations $\hat{\mathbf{k}} \cdot \mathbf{p}_i(t)$ of Eqs.\ \eqref{eq:tdiobsa} and \eqref{eq:tdiobs} to the GW phase:
\begin{eqnarray}
\phi_s(t + \hat{\mathbf{k}} \cdot \mathbf{p}_i) &\simeq&
\omega t + {\textstyle\frac{1}{2}}\dot{\omega} t^2 + (\omega + \dot{\omega} t) \, \hat{{\bf k}}\cdot[{\bf r} + {\sf O}_2(t) \cdot {\bf p}^L_i] \nonumber \\ &\simeq&
\underbrace{\omega t + {\textstyle\frac{1}{2}}\dot{\omega} t^2 + (\omega + \dot{\omega} t) \, R \cos\beta\cos(\Omega t + \eta_o - \lambda)}_{\phi(t)} + \underbrace{\omega \hat{{\bf k}} \cdot ({\sf O}_2(t) \cdot {\bf p}^L_i)}_{2 \omega L d_i(t)}
\label{fase} \label{eq:dopplershifting}
\end{eqnarray}
where $\phi(t)$ is the GW phase retarded to the position of the LISA guiding center, and where we have defined
\begin{equation}
d_i(t) \equiv \frac{\hat{{\bf k}} \cdot ({\sf O}_2(t) \cdot {\bf p}^L_i)}{2 L} =
   {\textstyle \frac{\sqrt{3}}{8}}\cos \beta \sin \gamma_{2i} - 
   {\textstyle \frac{1}{4}} \sin \beta \sin(\delta - \gamma_{2i}) +
   {\textstyle \frac{\sqrt{3}}{24}} \cos \beta \sin(2\delta - \gamma_{2i})
\label{di}
\end{equation}
(setting $\zeta = - \pi/6$, and with $\gamma_{2i} = \lambda - \eta_0 - \xi_0 + 2\sigma_i$). Equations \eqref{eq:tdiobsa} and \eqref{eq:tdiobs} contain also the projection factors $\hat{\bf k} \cdot \hat{{\bf n}}_i(t)$, which are given explicitly by
\begin{equation}      
c_i(t) \equiv - \hat{\bf k} \cdot [{\sf O}_2(t) \cdot \hat{{\bf n}}^L_i] = {\textstyle \frac{3}{4}}\cos\beta\sin\gamma_i  - {\textstyle \frac{\sqrt{3}}{2}}\sin\beta\sin(\delta - \gamma_i) + 
      {\textstyle \frac{1}{4}}\cos\beta\sin(2\delta - \gamma_i).
\label{ci}
\end{equation}
The functions $c_i(t)$ and $d_i(t)$ are related by $d_1 = (c_2 - c_3)/6$, $d_2 = (c_3 - c_1)/6$, and $d_3 = (c_1 - c_2)/6$.
Substituting the expressions \eqref{eq:tdiobs} [and similar ones] for the $y^\mathrm{GW}_{ij}$ into Eq.\ \eqref{eq:30} for $X^\mathrm{GW}_1$, we get after some algebra
\begin{equation}
X^\mathrm{GW}_1 (t) = 4 \, \omega L \, \sin(\omega L) \, \sin (2\omega L) \sum_{k=1}^{4} a^{(k)} X^{(k)}_1(t);
\label{X_1}
\end{equation}
the functions $X^{(k)}_1(t)$ are given by
\begin{eqnarray}
\left[\!\begin{array}{c}
X^{(1)}_1 \\
X^{(2)}_1
\end{array}\!\right]
&=&
- \left[\!\begin{array}{c}
u_2(t) \\
v_2(t)
\end{array}\!\right]\!
\bigl\{\mbox{sinc}\bigl[(1+c_2)x/2\bigr]\sin\bigl[\phi(t) - x d_2 - 7x/2\bigr]    
              + \mbox{sinc}\bigl[(1-c_2)x/2\bigr]\sin\bigl[\phi(t) - x d_2 -
              9x/2\bigr]\bigr\} \nonumber
\\ 
& & + 
\left[\!\begin{array}{c}
u_3(t) \\
v_3(t)
\end{array}\!\right]\!
\bigl\{\mbox{sinc}\bigl[(1+c_3)x/2\bigr]\sin\bigl[\phi(t) - x d_3 - 9x/2\bigr] 
              + \mbox{sinc}\bigl[(1-c_3)x/2\bigr]\sin\bigl[\phi(t) - x d_3 -
             7x/2\bigr]\bigr\}, \\ 
\left[\!\begin{array}{c}
X^{(3)}_1 \\
X^{(4)}_1
\end{array}\!\right]
&=&
\left[\!\begin{array}{c}
u_2(t) \\
v_2(t)
\end{array}\!\right]\!
\bigl\{\mbox{sinc}\bigl[(1+c_2)x/2\bigr]\cos\bigl[\phi(t) - x d_2 - 7x/2\bigr]    
              + \mbox{sinc}\bigl[(1-c_2)x/2\bigr]\cos\bigl[\phi(t) - x d_2 - 9x/2\bigr]\bigr\} \nonumber \\ 
& & -
\left[\!\begin{array}{c}
u_3(t) \\
v_3(t)
\end{array}\!\right]\!
\bigl\{\mbox{sinc}\bigl[(1+c_3)x/2\bigr]\cos\bigl[\phi(t) - x d_3 - 9x/2\bigr] 
              + \mbox{sinc}\bigl[(1-c_3)x/2\bigr]\cos\bigl[\phi(t) - x d_3 -
             7x/2\bigr]\bigr\},
\label{X_1last}
\end{eqnarray}
where $x = \omega L$ and ${\mathrm{sinc}\,\ldots = (\sin \ldots)/(\ldots)}$. 
The GW responses for $X_2$ and $X_3$ can be obtained by cyclical permutation of the spacecraft indices.

The GW response for $\alpha_1$ can be written in similar form:
\begin{equation}
\alpha^\mathrm{GW}_1 (t) = 2 \, \omega L \, \sin\bigl({\textstyle \frac{3}{2} \omega L}\bigr) \sum_{k=1}^{k=4} a^{(k)} \alpha^{(k)}_1(t),
\label{alpha_1}
\end{equation}
where
\begin{eqnarray}
\left[\!\begin{array}{c}
\alpha^{(1)}_1 \\
\alpha^{(2)}_1
\end{array}\!\right]
&=&
\left[\!\begin{array}{c}
u_1(t) \\ v_1(t)
\end{array}\!\right]
\bigl\{\mbox{sinc}\bigl[(1+c_1)x/2\bigr]\cos\bigl[\phi(t) - x d_1 - 3 x\bigr] 
- \mbox{sinc}\bigl[(1-c_1)x/2\bigr]\cos\bigl[\phi(t) - x
d_1 - 3 x\bigr]\bigr\} \nonumber \\ 
&& + 
\left[\!\begin{array}{c}
u_2(t) \\ v_2(t)
\end{array}\!\right]
\bigl\{\mbox{sinc}\bigl[(1+c_2)x/2\bigr]\cos\bigl[\phi(t) - x d_2 - 2 x\bigr]   
-    \mbox{sinc}\bigl[(1-c_2)x/2\bigr]\cos\bigl[\phi(t) - x
d_2 - 4 x\bigr]\bigr\} \label{alpha11} \\ 
&& + 
\left[\!\begin{array}{c}
u_3(t) \\ v_3(t)
\end{array}\!\right]
\bigl\{\mbox{sinc}\bigl[(1+c_3)x/2\bigr]\cos\bigl[\phi(t) - x d_3 - 4 x\bigr] 
-    \mbox{sinc}\bigl[(1-c_3)x/2\bigr]\cos\bigl[\phi(t) - x
d_3 - 2 x\bigr]\bigr\}, \nonumber \\ 
\left[\!\begin{array}{c}
\alpha^{(3)}_{1} \\ \alpha^{(4)}_{1}
\end{array}\!\right]
&=&
\left[\!\begin{array}{c}
u_1(t) \\ v_1(t)
\end{array}\!\right]
\bigl\{\mbox{sinc}\bigl[(1+c_1)x/2\bigr]\sin\bigl[\phi(t) - x d_1 - 3 x\bigr]  
- \mbox{sinc}\bigl[(1-c_1)x/2\bigr]\sin\bigl[\phi(t) - x
d_1 - 3 x\bigr]\bigr\} \nonumber \\
&& +
\left[\!\begin{array}{c}
u_2(t) \\ v_2(t)
\end{array}\!\right]
\bigl\{\mbox{sinc}\bigl[(1+c_2)x/2\bigr]\sin\bigl[\phi(t) - x d_2 - 2 x\bigr]    
- \mbox{sinc}\bigl[(1-c_2)x/2\bigr]\sin\bigl[\phi(t) - x d_2 - 4 x\bigr]\bigr\} \label{alpha14} \\ 
&& +
\left[\!\begin{array}{c}
u_3(t) \\ v_3(t)
\end{array}\!\right]
\bigl\{\mbox{sinc}\bigl[(1+c_3)x/2\bigr]\sin\bigl[\phi(t) - x d_3 - 4 x\bigr] 
- \mbox{sinc}\bigl[(1-c_3)x/2\bigr]\sin\bigl[\phi(t) - x
d_3 - 2 x\bigr]\bigr\}. \nonumber
\end{eqnarray}
The $\alpha_2$ and $\alpha_3$ combinations are again obtained by cyclical permutation of the spacecraft indices.

For the second-generation TDI observable $\zeta_1$ (see Ref.\ \cite{TEA03}; $\zeta_1$ is uniquely determined in the equal-armlength limit, unlike in the general case) we find 
\begin{equation}
\zeta^\mathrm{GW}_1 (t) = 2 \, \omega L \, \sin({\textstyle \frac{1}{2} \omega L}) \sum_{k=1}^{4} a^{(k)} \zeta^{(k)}_1(t),
\label{zeta_1}
\end{equation}
with
\begin{eqnarray}
\left[\!\begin{array}{c}
\zeta^{(1)}_1 \\ \zeta^{(2)}_1
\end{array}\!\right]
&=&
\left[\!\begin{array}{c}
u_1(t) \\ v_1(t)
\end{array}\!\right]
\bigl\{\mbox{sinc}\bigl[(1+c_1)x/2\bigr]\cos\bigl[\phi(t) - x d_1 - 3 x\bigr] 
- \mbox{sinc}\bigl[(1-c_1)x/2\bigr]\cos\bigl[\phi(t) - x
d_1 - 3 x\bigr]\bigr\} \nonumber \\ 
&& + 
\left[\!\begin{array}{c}
u_2(t) \\ v_2(t)
\end{array}\!\right]
\bigl\{\mbox{sinc}\bigl[(1+c_2)x/2\bigr]\cos\bigl[\phi(t) - x d_2 - 3 x\bigr]   
-    \mbox{sinc}\bigl[(1-c_2)x/2\bigr]\cos\bigl[\phi(t) - x
d_2 - 3 x\bigr]\bigr\} \label{zeta11} \\ 
&& + 
\left[\!\begin{array}{c}
u_3(t) \\ v_3(t)
\end{array}\!\right]
\bigl\{\mbox{sinc}\bigl[(1+c_3)x/2\bigr]\cos\bigl[\phi(t) - x d_3 - 3 x\bigr] 
-    \mbox{sinc}\bigl[(1-c_3)x/2\bigr]\cos\bigl[\phi(t) - x
d_3 - 3 x\bigr]\bigr\}, \nonumber \\ 
\left[\!\begin{array}{c}
\zeta^{(3)}_{1} \\ \zeta^{(4)}_{1}
\end{array}\!\right]
&=&
\left[\!\begin{array}{c}
u_1(t) \\ v_1(t)
\end{array}\!\right]
\bigl\{\mbox{sinc}\bigl[(1+c_1)x/2\bigr]\sin\bigl[\phi(t) - x d_1 - 3 x\bigr]  
- \mbox{sinc}\bigl[(1-c_1)x/2\bigr]\sin\bigl[\phi(t) - x
d_1 - 3 x\bigr]\bigr\} \nonumber \\
&& +
\left[\!\begin{array}{c}
u_2(t) \\ v_2(t)
\end{array}\!\right]
\bigl\{\mbox{sinc}\bigl[(1+c_2)x/2\bigr]\sin\bigl[\phi(t) - x d_2 - 3 x\bigr]    
- \mbox{sinc}\bigl[(1-c_2)x/2\bigr]\sin\bigl[\phi(t) - x 
d_2 - 3 x\bigr]\bigr\}  \label{zeta13} \\ 
&& +
\left[\!\begin{array}{c}
u_3(t) \\ v_3(t)
\end{array}\!\right]
\bigl\{\mbox{sinc}\bigl[(1+c_3)x/2\bigr]\sin\bigl[\phi(t) - x d_3 - 3 x\bigr] 
- \mbox{sinc}\bigl[(1-c_3)x/2\bigr]\sin\bigl[\phi(t) - x
d_3 - 3 x\bigr]\bigr\}. \nonumber
\end{eqnarray}
Finally, the \emph{optimal} TDI observables \cite{PTLA02}, which here we denote as
$\bar{A}$, $\bar{E}$ and $\bar{T}$ to
distinguish them from the optimal combinations $A$, $E$, and $T$
derived within first-generation TDI, are defined as linear combinations of $\alpha_1$, $\alpha_2$ and $\alpha_3$:
\begin{eqnarray}
\bar{A} &=& \frac{1}{\sqrt{2}}(\alpha_3 - \alpha_1), \nonumber \\
\bar{E} &=& \frac{1}{\sqrt{6}}(\alpha_1 - 2\alpha_2 +
\alpha_3), \label{eq:aetresp} \\
\bar{T} &=& \frac{1}{\sqrt{3}}(\alpha_1 +  \alpha_2 + \alpha_3). \nonumber
\end{eqnarray}
It is clear that $\bar{A}$, $\bar{E}$, and $\bar{T}$ are also optimal, in the sense discussed in Ref.\ \cite{PTLA02}: this is because they can be written as time-delayed combinations of the first-generation optimal TDI observables, such as $\bar{A} = A(t) - A(t - 3L)$, $\bar{E} = E(t) - E(t - 3L)$, and $\bar{T} = T(t) - T(t - 3L)$; since by construction the noises
that enter $A$, $E$, and $T$ are uncorrelated, it follows that the
noises that enter $\bar{A}$, $\bar{E}$, and $\bar{T}$ are also uncorrelated, making these observables optimal. 

We recall that in the high-frequency part of the LISA band (i.e., for frequencies equal to or larger than $5$ mHz), there exist three independent
TDI GW observables (such as $\bar{A}$, $\bar{E}$, and $\bar{T}$, or $X_1$, $X_2$, and $X_3$). However, for frequencies smaller than $5$ mHz, there are essentially only two independent observables: this is especially obvious if we reason in terms of the optimal combinations, where we observe that
for low frequencies the GW signal response of $\bar{T}$ declines much faster than the responses of $\bar{A}$ and $\bar{E}$ \cite{TAE01,PTLA02}.

\subsection{TDI responses in the long-wavelength limit}

The long-wavelength (LW) approximation to the GW responses is obtained by taking the leading-order terms of the generic expressions in the limit of $\omega L \rightarrow 0$. For instance, for $X_1$ [Eqs.\ \eqref{X_1}--\eqref{X_1last}], we get
\begin{equation}
X^\mathrm{GW}_{1,\mathrm{LW}} \simeq 16 (\omega L)^3 \left\{
\left[ u_3(t) - u_2(t) \right] \left[ a^{(1)} \sin \phi(t) - a^{(3)} \cos \phi(t) \right]+
\left[ v_3(t) - v_2(t) \right] \left[ a^{(2)} \sin \phi(t) - a^{(4)} \cos \phi(t) \right]
\right\},
\label{eq:x1lw}
\end{equation}
with $a^{(k)}$ given by Eqs.\ \eqref{eq:ampone}--\eqref{Aamp}, and $u_i(t)$, $v_i(t)$ by Eqs.\ \eqref{ak}, \eqref{bk}. The LW responses for $X_2$ and $X_3$ can be obtained by cyclical permutation of the indices. Adopting the notation of Ref.\ \cite{AET99}, we find also that
\begin{equation}
X^\mathrm{GW}_{1,\mathrm{LW}} (t) \simeq 8 L^3 \left[(\hat{\bf n}^L_3)' \cdot \dddot{{\sf H}}^L(t) \cdot \hat{\bf n}^L_3 
- (\hat{\bf n}^L_2)' \cdot \dddot{{\sf H}}^L(t) \cdot \hat{\bf n}^L_2\right],
\end{equation}
where ``$^\dddot{}$'' denotes the third time derivative,
$\hat{\bf n}^L_i$ is given by Eq.\ \eqref{eq:nlisa},
and ${\sf H}^L(t)$ is given by Eq.\ \eqref{HT}.

The GW responses of the Sagnac observables $\alpha^\mathrm{GW}_{i,\mathrm{LW}}$ are equal simply to $\frac{3}{8} X^\mathrm{GW}_{i,\mathrm{LW}}$. From Eqs.\ \eqref{eq:aetresp} we then get the LW GW responses $\bar{A}_\mathrm{LW}^\mathrm{GW}$, $\bar{E}_\mathrm{LW}^\mathrm{GW}$, and $\bar{T}_\mathrm{LW}^\mathrm{GW}$:
\begin{align}
\bar{A}^\mathrm{GW}_{\mathrm{LW}} &\simeq 3 \sqrt{2} \, (\omega L)^3 \left\{
\left[ 2 u_2(t) - u_1(t) - u_3(t) \right] \left[ a^{(1)} \sin \phi(t) - a^{(3)} \cos \phi(t) \right] \right. \nonumber \\
& \phantom{\simeq 3 \sqrt{2} \, (\omega L)^3 \left\{ \left[ 2 u_2(t) - u_1(t) - u_3(t) \right] \left[ a^{(1)} \sin \right. \right.} + \left. \left[ 2 v_2(t) - v_1(t) - v_3(t) \right] \left[ a^{(2)} \sin \phi(t) - a^{(4)} \cos \phi(t) \right] \right\}, \label{eq:mylwa} \\
\bar{E}^\mathrm{GW}_{\mathrm{LW}} &\simeq 3 \sqrt{6} \, (\omega L)^3 \left\{
\left[ u_3(t) - u_1(t) \right] \left[ a^{(1)} \sin \phi(t) - a^{(3)} \cos \phi(t) \right] + \left[ v_3(t) - v_1(t) \right] \left[ a^{(2)} \sin \phi(t) - a^{(4)} \cos \phi(t) \right] \right\}, \label{eq:mylwe} \\
\bar{T}^\mathrm{GW}_{\mathrm{LW}} &\simeq O[(\omega L)^4]. \label{eq:mylwt}
\end{align}

\section{Noise spectral density}

The spectral density of noise for the first-generation TDI observables $X$, $Y$, $Z$, $\alpha$, $\beta$, $\gamma$, $A$, $E$, and $T$ is given in Refs.\ \cite{ETA00,PTLA02} in the case of an equilateral LISA array, assuming that the noises appearing in all the proof masses and optical paths are uncorrelated.
The finite-difference relations between first- and second-generation TDI observables [such as $X_1 (t) = X(t) - X(t - 4L)$, $\alpha_1 (t) = \alpha(t) - \alpha(t - 3L)$] imply simple modifications to the first-generation noise densities: for instance,
\begin{eqnarray}
S_{X_1} (\omega) = 4 \, \sin^2(2 \omega L) \, S_{X} (\omega),
\label{SX1} \\
S_{\alpha_1} (\omega) = 4 \, \sin^2(3 \omega L/2) \, S_{\alpha} (\omega);
\label{Sal1}
\end{eqnarray}
inserting the expression of $S_X = S_Y = S_Z$ from Ref.\ \cite{ETA00} into Eq.\ \eqref{SX1} yields
\begin{equation}
S_{X_1} = S_{X_2} = S_{X_3} = 64 \sin^2\,(\omega L) \sin^2\,(2 \omega L) \bigl[2 (1+\cos^2 \omega L) S^\mathrm{pm} +  S^\mathrm{op}\bigr],
\label{eq:spec}
\end{equation}
where $S^\mathrm{pm} = 2.54 \times 10^{-48} f^{-2} \, \mathrm{Hz}^{-1}$ and
$S^\mathrm{op} = 1.76 \times 10^{-37} f^2 \, \mathrm{Hz}^{-1}$ are
the fractional-frequency-fluctuation spectral densities of proof-mass noise and optical-path noise respectively \cite{ETA00}.
These values correspond to an rms single-proof-mass acceleration noise of $3 \times 10^{-15}$ m sec$^{-2}$ Hz$^{-1/2}$, and to an rms aggregate optical-path noise $20 \times 10^{-12}$ m Hz$^{-1/2}$, as quoted in the LISA Pre-Phase A Study \cite{PPA98}. For the other TDI observables we find
\begin{equation}
S_{\alpha_1} = S_{\alpha_2} = S_{\alpha_3} = 8 \sin^2(3 \omega L/2) \bigl(\bigl[4 \sin^2(3 \omega L/2) + 8\,\sin^2(\omega L/2)\bigr] S^\mathrm{pm} + 3 \, S^\mathrm{op}\bigr),
\end{equation}
\begin{equation}
S_{\bar{A}} = S_{\bar{E}} = 32 \sin^2(\omega L/2) \sin^2(3 \omega L/2) \bigl(\bigl[6 + 4 \cos (\omega L) + 2 \cos(2\omega L)\bigr] S^\mathrm{pm} + \bigl[2 + \cos (\omega L)\bigr]\,S^\mathrm{op}\bigr),
\end{equation}
\begin{equation}
S_{\bar{T}} = 8 \bigl[1 + 2 \cos (\omega L)\bigr]^2 \sin^2(3 \omega L/2)
\bigl[4\sin^2(\omega L/2) S^\mathrm{pm} +  S^\mathrm{op} \bigr]. 
\end{equation}
All the noise spectra are shown in Fig.\ 5. In the long-wavelength approximation, the noise expressions simplify to
\begin{equation}
S^\mathrm{LW}_{X_1} = S^\mathrm{LW}_{X_2} = S^\mathrm{LW}_{X_3}  \simeq  256 (\omega L)^2 [4 (\omega L)^2 S^\mathrm{pm}  + (\omega L)^2 S^\mathrm{op}],
\end{equation}
\begin{equation}
S^\mathrm{LW}_{\alpha_1} = S^\mathrm{LW}_{\alpha_2} = S^\mathrm{LW}_{\alpha_3}  \simeq  18 (\omega L)^2 [11 (\omega L)^2 S^\mathrm{pm}  +  3 S^\mathrm{op}],
\end{equation}
\begin{equation}
S^\mathrm{LW}_{\bar{A}} = S^\mathrm{LW}_{\bar{E}}  \simeq  54 (\omega L)^2 [4 (\omega L)^2 S^\mathrm{pm}  + (\omega L)^2 S^\mathrm{op}],
\end{equation}
\begin{equation}
S^\mathrm{LW}_{\bar{T}}  \simeq  162 (\omega L)^2 [(\omega L)^2 S^\mathrm{pm}  +  S^\mathrm{op}].
\end{equation}
\begin{figure}
\label{Fig:noi}
  \includegraphics[width=11.5cm]{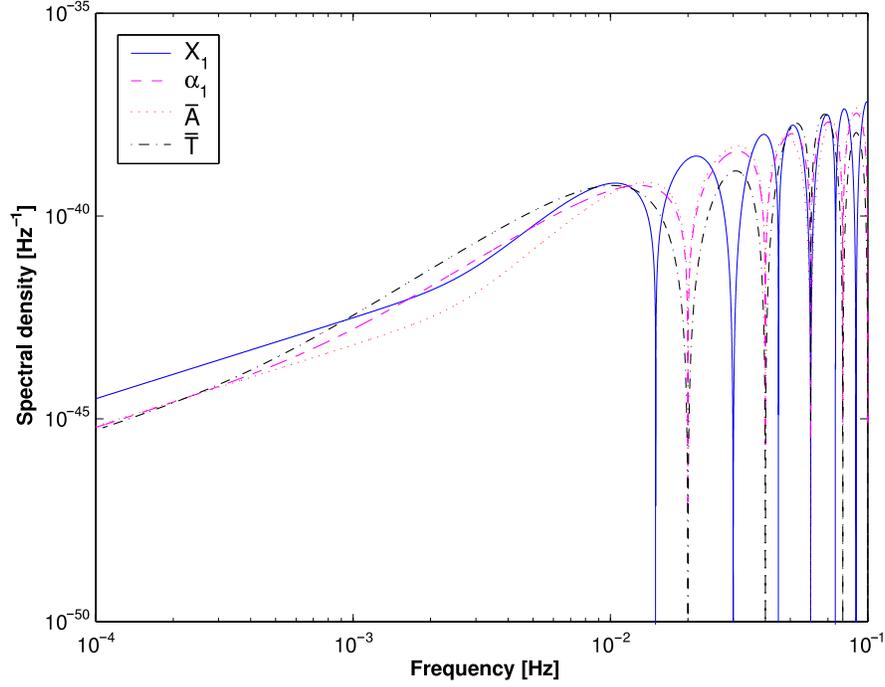}
  \caption{Spectral densities of noise for the second-generation TDI observables $X_1$ (continuous), $\alpha_1$ (dashed), $\bar{A}$ (dotted), and $\bar{T}$ (dash-dotted).}
\end{figure}

\section{Optimal filtering of the LISA data}

In this section we develop a \emph{maximum-likelihood} (ML) formalism to detect GW signals from moderately chirping binaries and to estimate their parameters, by analyzing the time series of the TDI observables. ML detection is based on maximizing the \emph{likelihood ratio} $\Lambda(\theta^i)$ over the source parameters $\theta^i$; this ratio is proportional to the probability that the observed detector output could have been produced by a GW source with parameters $\theta^i$, plus instrument noise. The magnitude of the maximum indicates the probability that a signal is indeed present, while its location indicates the most likely parameters (the ML \emph{parameter estimators}). Under the assumption of Gaussian, stationary, additive noise, $\log \Lambda(\theta^i)$ is computed by correlating the detector output, $x(t)$, with the expected GW detector response $h_{\theta^i}(t)$, while weighting the correlation in the frequency domain by the inverse spectral density of instrument noise, $S^{-1}_n(f)$. The family of GW responses $\{h_{\theta^i}(t)\}$, divided (in the frequency domain) by $S_n(f)$ to incorporate the noise weighting, are known as \emph{optimal filters}.

In Sec.\ \ref{sec:ml} we describe the computation of $\Lambda$ and of the ML parameter estimators for the optimal filters derived from the GW responses of Sec.\ II, and we show how to maximize $\Lambda$ \emph{algebraically} over the four source amplitudes $a^{(k)}$. 
The amplitude-maximized $\log \Lambda$ (known as $\mathcal{F}$) is then
used as a \emph{detection statistic} to search for the most likely GW source, by maximizing it over the remaining source parameters [here denoted by $\xi^\mu$; thus, $\theta^i \equiv (a^{(k)},\xi^\mu)$]. In Sec.\ \ref{sec:fdist} we study the statistical distribution of $\mathcal{F}(\xi^\mu)$ in the absence (or presence) of a GW signal of parameters $\xi^\mu$; this distribution determines the statistical significance of observing a certain value of $\mathcal{F}$, \emph{for a fixed} $\xi^\mu$. In Sec.\ \ref{sec:false} we study the statistical significance of measuring a certain value of the \emph{completely maximized} statistic $\max_{\xi^\mu} \mathcal{F}(\xi^\mu)$, which leads to the total \emph{false-alarm probability} for a GW search over a range of intrinsic parameters.  See Helstr\"om \cite{H68} for an extended discussion of ML detection and parameter estimation.

\subsection{Maximum-likelihood search method}
\label{sec:ml}

As discussed in Secs.\ I and II, the LISA Doppler measurements can be recombined into the laser-noise and optical-bench--noise free TDI observables, all of which can be obtained as time-delayed combinations of three generators \cite{AET99,DNV}. Thus, in the following we denote the TDI data as the three-vector $\mathbf{x}(t)$ (we shall very shortly specify a convenient vector basis). In the case of additive noise, we can write $\mathbf{x}(t) = \mathbf{n}(t) + \mathbf{h}(t)$, where $\mathbf{n}(t)$ represents detector noise and $\mathbf{h}(t)$ the GW response. Idealizing $\mathbf{n}(t)$ as a zero-mean, Gaussian, stationary, continuous random process, we have \cite{H68}
\begin{equation}
\label{eq:L}
\log\Lambda = (\bf{x}|\bf{h}) - {\textstyle\frac{1}{2}}(\bf{h}|\bf{h}),
\end{equation}
where the scalar product $(\ldots|\ldots)$ is defined by
\begin{equation}
\label{eq:SP}
(\mathbf{x}|\mathbf{y}) =
4 \, \mathrm{Re} \int^{\infty}_{0}\tilde{\bf{x}}^\dagger \cdot \tilde{\bf{S}}^{-1}_n \cdot \tilde{\bf{y}}
\,df;
\end{equation}
here the dagger denotes transposition and complex-conjugation, 
the tilde denotes the Fourier transform, 
and $\tilde{\bf{S}}$ denotes the one-sided cross spectral density
matrix of detector noise, defined by the expectation value
\begin{equation}
\label{eq:CSD}
{\cal E} [\tilde{\bf{n}}(f)\tilde{\bf{n}}^{\dagger}(f')] = {\textstyle \frac{1}{2}}\delta(f - f')\tilde{\bf{S}}_n(f).
\end{equation}

The larger the signal with respect to the noise, the higher the probability that a ML search (performed with the appropriate optimal filter) will yield a statistically significant detection, and the better the accuracy of the ML parameter estimators. The accuracy of estimation is also better for the parameters on which the signal is strongly dependent.
Signal strength is characterized by the \emph{optimal signal-to-noise ratio} (optimal S/N), 
\begin{equation}
\label{eq:d}
\rho^2 = (\bf{h}|\bf{h});
\end{equation}
while the dependence of the instrument response on the parameters is characterized by the Fisher information matrix,
\begin{equation}
\label{eq:GA}
\Gamma_{ij} = 
\bigg(\frac{\partial {\bf h}}{\partial\theta^i}\bigg|\frac{\partial{\bf h}}{\partial \theta^j}\bigg).
\end{equation}
By the Cram\'er--Rao inequality \cite{H68}, the
diagonal elements of $\Gamma^{-1}_{ij}$ provide lower bounds on the variance of any unbiased estimators of the $\theta^i$. In fact, the matrix $\Gamma^{-1}_{ij}$ is often called the \emph{covariance matrix}, because in the limit of high S/N the ML estimators become unbiased, and their distribution tends to a jointly Gaussian distribution with covariance matrix equal to $\Gamma^{-1}_{ij}$.

The optimal TDI observables \cite{PTLA02} are obtained by diagonalizing the cross spectrum $\tilde{\bf S}_n$; it turns out that the eigenvectors $\bar{A}$, $\bar{E}$, and $\bar{T}$ are independent of frequency. The new observables are the linear combinations of the Sagnac observables $\alpha_1, \alpha_2, \alpha_3$ given by Eq.\ \eqref{eq:aetresp}, and by definition their noises are uncorrelated.
With reference to Eq.\ \eqref{eq:spec}, we define $S_{\bar{A}\bar{A}}(f) = S_{\bar{E}\bar{E}}(f) \equiv S_{\bar{A}}(\omega \equiv 2 \pi f)$ and $S_{\bar{T}\bar{T}}(f) \equiv S_{\bar{T}}(\omega \equiv 2 \pi f)$. It is convenient to use the optimal observables $\bar{A}$, $\bar{E}$, and $\bar{T}$ as a basis for the LISA TDI observables, setting
\begin{equation}
\mathbf{x}(t) =
\left[\begin{array}{c}
\bar{A}(t) \\ \bar{E}(t) \\ \bar{T}(t)
\end{array}\right], \quad
\mathbf{h}(t) =
\left[\begin{array}{c}
\bar{A}^\mathrm{GW}(t) \\ \bar{E}^\mathrm{GW}(t) \\ \bar{T}^\mathrm{GW}(t)
\end{array}\right];
\end{equation}
the GW responses $\bar{A}^\mathrm{GW}$, $\bar{E}^\mathrm{GW}$, and $\bar{T}^\mathrm{GW}$ are given in Sec.\ \ref{sec:tdiresp} for the case of moderately chirping binaries. For these sources, $S_{\bar{A}}(\omega)$ and $S_{\bar{T}}(\omega)$ are approximately constant over the signal bandwidth, so we can expand Eq.\ \eqref{eq:L} as
\begin{eqnarray}
\log\Lambda & \cong &
T_0 \left[ \big(\bar{A}\big|\big|\bar{A}^\mathrm{GW}\big) - {\textstyle\frac{1}{2}}\big(\bar{A}^\mathrm{GW}\big|\big|\bar{A}^\mathrm{GW}\big) + 
 \big(\bar{E}\big|\big|\bar{E}^\mathrm{GW}\big) - {\textstyle\frac{1}{2}}\big(\bar{E}^\mathrm{GW}\big|\big|\bar{E}^\mathrm{GW}\big) \right] / S_{\bar{A}}(\omega) 
\nonumber \\
& & + \; 
T_0 \left[ \big(\bar{T}\big|\big|\bar{T}^\mathrm{GW}\big) - {\textstyle\frac{1}{2}}\big(\bar{T}^\mathrm{GW}\big|\big|\bar{T}^\mathrm{GW}\big) \right] / S_{\bar{T}}(\omega), \label{eq:likelihood}
\end{eqnarray}
where $T_0$ is the time of observation, and where we have introduced the time-domain scalar product
\begin{equation}
(B||C) \equiv (2/T_0)\int^{T_0}_{0} B(t) C(t)\,dt.
\end{equation}
Given the linear dependence \cite{PTLA02} of $\bar{A}$, $\bar{E}$, and $\bar{T}$ on $\alpha_1$, $\alpha_2$, and $\alpha_3$, from Eq.\ \eqref{alpha_1} it follows that
\begin{equation}
\left[\begin{array}{c}
\bar{A}^\mathrm{GW}(t) \\ \bar{E}^\mathrm{GW}(t) \\ \bar{T}^\mathrm{GW}(t)
\end{array}\right] = 2 x \sin ({\textstyle \frac{3}{2}} x)
\sum_{k=1}^{4} a^{(k)} \!
\left[\begin{array}{c}
\bar{A}^{(k)}(t) \\ \bar{E}^{(k)}(t) \\ \bar{T}^{(k)}(t)
\end{array}\right] = 2 x \sin ({\textstyle \frac{3}{2}} x)
\sum_{k=1}^{4} a^{(k)} \!
\left[\begin{array}{c}
{\textstyle \frac{1}{\sqrt{2}}} \left( \alpha^{(k)}_3 - \alpha^{(k)}_1 \right) \\
{\textstyle \frac{1}{\sqrt{6}}} \left( \alpha^{(k)}_1 - 2 \alpha^{(k)}_2 + \alpha^{(k)}_3 \right) \\
{\textstyle \frac{1}{\sqrt{3}}} \left( \alpha^{(k)}_1 + \alpha^{(k)}_2 + \alpha^{(k)}_3 \right)
\end{array}\right],
\label{eq:aetdecomp}
\end{equation}
where once again $x = \omega L$, the amplitudes $a^{(k)}$ are given by Eqs.\ \eqref{eq:ampone}--\eqref{Aamp}, and the functions $\alpha_i^{(k)}$ are given by Eqs.\ \eqref{alpha11} and \eqref{alpha14}, and by similar equations obtained by cyclical permutation of the indices. 
Note that the component functions $\bar{A}^{(k)}(t)$, $\bar{E}^{(k)}(t)$, and $\bar{T}^{(k)}(t)$ do not depend on the amplitudes $a^{(k)}$ (or equivalently, on $h_0^+$, $h_0^\times$, $\phi_0$, and $\psi$); they do however depend on the remaining (intrinsic) source parameters, $\omega$, $\dot{\omega}$, $\beta$, and $\lambda$. 

The ML parameter estimators $\hat{\theta}^i$ are found by maximizing $\log \Lambda$ with respect to the source parameters $\theta^i$: that is, by solving
\begin{equation}
\frac{\partial\log\Lambda}{\partial \theta^i} = 0.
\end{equation}
For the $a^{(k)}$ this is accomplished easily by solving the linear system
\begin{equation}
\sum_{k=1}^4 M^{(l)(k)} a^{(k)} = N^{(l)},\quad l=1,\ldots,4,
\label{eq:linsys}
\end{equation}
where
\begin{equation}
N^{(l)} = 2 x \sin ({\textstyle \frac{3}{2}} x) T_0 \, \left[
  (\bar{A}||\bar{A}^{(l)}) / S_{\bar{A}}(\omega)
+ (\bar{E}||\bar{E}^{(l)}) / S_{\bar{A}}(\omega)
+ (\bar{T}||\bar{T}^{(l)}) / S_{\bar{T}}(\omega) \right],
\label{eq:ndef}
\end{equation}
and where $M^{(l)(k)}$ is the $4 \times 4$ matrix with components
\begin{equation}
\label{MLE}
M^{(l)(k)} = 4 x^2 \sin^2 ({\textstyle \frac{3}{2}} x) T_0 \, \left[
  (\bar{A}^{(l)}||\bar{A}^{(k)}) / S_{\bar{A}}(\omega)
+ (\bar{E}^{(l)}||\bar{E}^{(k)}) / S_{\bar{A}}(\omega)
+ (\bar{T}^{(l)}||\bar{T}^{(k)}) / S_{\bar{T}}(\omega) \right].
\end{equation}
The solution of Eq.\ \eqref{eq:linsys} is simplified by noticing that
the component functions $\bar{A}^{(k)}(t)$, $\bar{E}^{(k)}(t)$, and $\bar{T}^{(k)}(t)$ consist of simple sines and cosines with period $\sim 2\pi/\omega$, modulated by the slowly changing functions $u_i(t)$ and $v_i(t)$ (with periods that are multiples of 1 yr).
By the approximate orthogonality of sine and cosine terms, for $T_0 \simeq 1$ yr the scalar products $(\bar{A}^{(k)}||\bar{A}^{(l)})$ can be approximated as
\begin{equation}
(\bar{A}^{(1)}||\bar{A}^{(3)}) \simeq
(\bar{A}^{(2)}||\bar{A}^{(4)}) \simeq 0,
\end{equation}
and
\begin{eqnarray}
(\bar{A}^{(1)}||\bar{A}^{(1)}) &\simeq& (\bar{A}^{(3)}||\bar{A}^{(3)}) \equiv {\textstyle\frac{1}{2}} U_{\bar{A}}, \\
(\bar{A}^{(2)}||\bar{A}^{(2)}) &\simeq& (\bar{A}^{(4)}||\bar{A}^{(4)}) \equiv {\textstyle\frac{1}{2}} V_{\bar{A}}, \\ 
(\bar{A}^{(1)}||\bar{A}^{(2)}) &\simeq& (\bar{A}^{(3)}||\bar{A}^{(4)}) \equiv {\textstyle\frac{1}{2}} Q_{\bar{A}}, \\
(\bar{A}^{(1)}||\bar{A}^{(4)}) &\simeq&-(\bar{A}^{(2)}||\bar{A}^{(3)}) \equiv {\textstyle\frac{1}{2}} P_{\bar{A}},
\end{eqnarray}
with similar expressions for the $\bar{E}^{(k)}$ and $\bar{T}^{(k)}$.
The matrix $M^{(l)(k)}$ then simplifies to
\begin{eqnarray}
M^{(l)(k)} = \frac{T_0}{2}
\left(\begin{array}{cccc}
 U & Q &  0 & P \\
 Q & V & -P & 0 \\
 0 & -P &  U & Q \\
 P & 0 &  Q & V 
\end{array}\right),
\label{M}
\end{eqnarray} 
where the elements $U$, $V$, $Q$, and $P$ are given by
\begin{eqnarray}
\label{eq:U}
U &=& 4 x^2 \sin^2 ({\textstyle \frac{3}{2}} x) \left[
U_{\bar{A}} / S_{\bar{A}}(\omega) +
U_{\bar{E}} / S_{\bar{A}}(\omega) +
U_{\bar{T}} / S_{\bar{T}}(\omega) \right], \\
\label{eq:V}
V &=& 4 x^2 \sin^2 ({\textstyle \frac{3}{2}} x) \left[
V_{\bar{A}} / S_{\bar{A}}(\omega) +
V_{\bar{E}} / S_{\bar{A}}(\omega) +
V_{\bar{T}} / S_{\bar{T}}(\omega) \right], \\
\label{eq:Q}
Q &=& 4 x^2 \sin^2 ({\textstyle \frac{3}{2}} x) \left[
Q_{\bar{A}} / S_{\bar{A}}(\omega) +
Q_{\bar{E}} / S_{\bar{A}}(\omega) +
Q_{\bar{T}} / S_{\bar{T}}(\omega) \right], \\
\label{eq:P}    
P &=& 4 x^2 \sin^2 ({\textstyle \frac{3}{2}} x) \left[
P_{\bar{A}} / S_{\bar{A}}(\omega) +
P_{\bar{E}} / S_{\bar{A}}(\omega) +
P_{\bar{T}} / S_{\bar{T}}(\omega) \right].
\end{eqnarray}
Then the analytic expressions for the maximum likelihood estimators 
$\hat{a}^{(k)}$ of the amplitudes $a^{(k)}$ are given by
\begin{equation}
\label{amle}
\left(\begin{array}{c}
\hat{a}^{(1)} \\ \hat{a}^{(2)} \\ \hat{a}^{(3)} \\ \hat{a}^{(4)}
\end{array}\right) = \frac{2}{T_0 \Delta}
\left(\begin{array}{cccc}
 V & -Q &  0  & -P \\
-Q &  U &  P  &  0 \\
 0 &  P &  V  & -Q \\
-P &  0 & -Q  &  U
\end{array}\right) \cdot
\left(\begin{array}{c}
N^{(1)} \\ N^{(2)} \\ N^{(3)} \\ N^{(4)}
\end{array}\right),
\end{equation}
where $\Delta = U V  - Q^2 - P^2$.

Substituting the ML amplitude estimators $\hat{a}^{(k)}$ in the likelihood function $\Lambda$ yields the \emph{reduced likelihood function} $\Lambda_r$.
The logarithm of $\Lambda_r$ is known as the $\mathcal{F}$ \emph{statistic}; using Eqs.\ \eqref{eq:likelihood}, \eqref{eq:aetdecomp}, \eqref{eq:ndef}, and \eqref{amle}, we find
\begin{multline}
\mathcal{F} = {\textstyle \frac{1}{2}} \sum_{l=1}^{4} \sum_{k=1}^{4} 
\big(M^{-1}\big)^{(l)(k)} N^{(l)} N^{(k)} 
= (T_0 \Delta)^{-1} \left\{
V \left[ (N^{(1)})^2 + (N^{(3)})^2 \right] +  U  \left[ (N^{(2)})^2 + (N^{(4)})^2 \right] 
\right. \\ \left. - 2 Q \left[ N^{(1)} N^{(2)} + N^{(3)} N^{(4)} \right] - 2 P \left[ N^{(1)} N^{(4)} - N^{(2)} N^{(3)} \right]
\right\}.
\label{eq:fdef}
\end{multline}
We adopt $\mathcal{F}$ as the detection statistic of our proposed search scheme. The statistic $\mathcal{F}$ is already maximized over the amplitudes $a^{(k)}$, which are known in this context as \emph{extrinsic parameters}. By contrast, the ML estimators of the remaining (\emph{intrinsic}) source parameters ($\omega$, $\dot{\omega}$, $\beta$, and $\lambda$) are found by maximizing $\mathcal{F}$. In practice, this is done by correlating the detector output with a \emph{bank} of optimal filters precomputed for many values of the intrinsic parameters. 

Introducing the complex quantities
\begin{eqnarray}
\label{eq:CA}
a^{(u)} &=& a^{(1)} + i a^{(3)}, \\
\label{eq:CB}
a^{(v)} &=& a^{(2)} + i a^{(4)}, \\
\label{eq:CQ}
W   &=& Q + i P, \\
\label{eq:CFa}
N^{(u)} &=& N^{(1)} + i N^{(3)}, \\
\label{eq:CFb}
N^{(v)} &=& N^{(2)} + i N^{(4)},
\end{eqnarray}
we can write the ML amplitude estimators and the ${\mathcal F}$ statistic in the compact form
\begin{eqnarray}
\label{eq:mld1}
\hat{a}^{(u)} &=&  2 (T_0 \Delta)^{-1} \bigl[V N^{(u)} - W^*  N^{(v)} \bigr], \\
\label{eq:mld2}
\hat{a}^{(v)} &=&  2 (T_0 \Delta)^{-1} \bigl[U N^{(v)} - W \, N^{(u)} \bigr],
\end{eqnarray}
[where $\Delta = U V - |W|^2$] and
\begin{equation}
\label{eq:mld3}
{\mathcal F} = (T_0 \Delta)^{-1} \left\{ V \bigl|N^{(u)}\bigr|^2 + U \bigl|N^{(v)}\bigr|^2 - 2 \, \mathrm{Re} \left[ W \, N^{(u)} (N^{(v)})^* \right] \right\}.
\end{equation}
In Sec.\ \ref{ssec:alg} we shall see that this expression is very suitable for numerical implementation. Equations \eqref{eq:mld1}--\eqref{eq:mld3} summarize the proposed ML data-analysis scheme, which uses all the available LISA data. Similar expressions hold if we analyze a single interferometric combination, such as $X_1$. 
In Appendix \ref{app:c} we describe a useful complex representation of the GW TDI responses that simplifies the integrals involved in the computation of ${\mathcal F}$ and of the ML amplitude estimators.

In the LW approximation, Eqs.\ \eqref{eq:mld1}--\eqref{eq:mld3} simplify somewhat:
using the $\bar{A}^\mathrm{GW}_\mathrm{LW}(t)$ and $\bar{E}^\mathrm{GW}_\mathrm{LW}(t)$ of Eqs.\ \eqref{eq:mylwa} and \eqref{eq:mylwe} [and remembering that $\bar{T}^\mathrm{GW}_\mathrm{LW}(t) \simeq 0$], we go through with our formalism in parallel with Eqs.\ \eqref{eq:ndef}--\eqref{M}, and find that $P_\mathrm{LW} \simeq 0$, so $W_\mathrm{LW}$ is real. The complex variables $N^{(u)}$ and $N^{(v)}$ are given by the integrals
\begin{eqnarray}
N_\mathrm{LW}^{(u)} &=& -2 i \frac{(\omega L)^3}{S^\mathrm{LW}_{\bar{A}}(\omega)}
\int_0^{T_0}
\left\{
3 \sqrt{2} \big[ 2 u_2(t) - u_1(t) - u_3(t) \big] \bar{A}(t) +
3 \sqrt{6} \big[ u_3(t) - u_1(t) \big] \bar{E}(t)
\right\} e^{i \phi(t)} \, dt, \\
N_\mathrm{LW}^{(v)} &=& -2 i \frac{(\omega L)^3}{S^\mathrm{LW}_{\bar{A}}(\omega)}
\int_0^{T_0}
\left\{
3 \sqrt{2} \big[ 2 v_2(t) - v_1(t) - v_3(t) \big] \bar{A}(t) +
3 \sqrt{6} \big[ v_3(t) - v_1(t) \big] \bar{E}(t)
\right\} e^{i \phi(t)} \, dt.
\end{eqnarray}
Analogous LW expressions hold for a single TDI observable, such as $X_{1}$.

\subsection{Distribution of the $\mathcal{F}$ statistic}
\label{sec:fdist}

Crucial to a search scheme based on comparing the ML statistic $\mathcal{F}$ with a predefined threshold is the determination of the false-alarm probability $P_F$ (which determines how often $\mathcal{F}$ will exceed the threshold in presence of noise alone) and of the detection probability $P_D$ (which determines how often $\mathcal{F}$ will exceed the threshold when a signal is present, resulting in correct detection). In this section we compute the probabilities $P_F$ and $P_D$ for the correlation of detector data against a single optimal filter (i.e., for fixed values of the intrinsic parameters).

Under the assumption of zero-mean Gaussian noise, the weighted correlations $N^{(k)}$ [Eq.\ \eqref{eq:ndef}] are Gaussian random variables; since $\mathcal{F}$ is a quadratic form in the $N^{(k)}$ [see Eq.\ \eqref{eq:fdef}], it must follow the $\chi^2$ distribution. Following Sec.\ III B of Ref.\ \cite{JK00},
we can diagonalize the quadratic form to find that, in the absence of the signal,
$2 \mathcal{F}$ follows the $\chi^2$ distribution with $n = 4$ degrees of freedom \cite{statnote}.
In presence of the signal, $2{\mathcal F}$ follows a noncentral $\chi^2$ distribution with $n = 4$ degrees of freedom and with noncentrality parameter $\kappa$ equal to the optimal $(\mathrm{S}/\mathrm{N})^2 = \rho^2$ [see Eq.\ \eqref{eq:d}]. For instance, if we use $\bar{A}$, $\bar{E}$, $\bar{T}$,
\begin{equation}
\kappa = \rho^2 =
\left(\bar{A}^\mathrm{GW}|\bar{A}^\mathrm{GW}\right) +
\left(\bar{E}^\mathrm{GW}|\bar{E}^\mathrm{GW}\right) +
\left(\bar{T}^\mathrm{GW}|\bar{T}^\mathrm{GW}\right)
\end{equation}
(which agrees with the result derived in Ref.\ \cite{PTLA02}), while if we use only $X_1$,
\begin{equation}
\kappa = \rho^2 = \left(X_1^\mathrm{GW}|X_1^\mathrm{GW}\right).
\end{equation}
The $\chi^2$ probability density function is
\begin{equation}
\label{p0}
p_0({\mathcal F}) = 
\frac{{\mathcal F}^{n/2-1}}{(n/2 -1)!}\exp(-{\mathcal F})
\end{equation}
for $\kappa = 0$, or
\begin{equation}
\label{p1}
p_1(\rho;\mathcal F) = 
\frac{(2{\mathcal F})^{(n/2 -1)/2}}{\rho^{n/2-1}} 
I_{n/2-1}\!\Bigl(\rho\sqrt{2 {\mathcal F}}\Bigr)
\exp \Bigl(-{\mathcal F}-{\textstyle\frac{1}{2}}\rho^2\Bigr)
\end{equation}
for $\kappa = \rho^2$, where $I_{n/2-1}$ is the $(n/2-1)$th-order modified Bessel function of the first kind. Thus, the false-alarm probability for a threshold $\mathcal{F}_0$ is
\begin{equation}
\label{PF}
P_F(\Fo) = \int_{\Fo}^\infty p_0(\F)\,d\F
= \exp(-\Fo) \sum^{n/2-1}_{k=0} \Fo^k / k!
\end{equation}
(for even $n$; for odd $n$ the result involves the error function) while the detection probability, in the presence of a $\mathrm{S}/\mathrm{N} = \rho$ signal (using the correct optimal filter), is
\begin{equation}
\label{PD}
P_D(\rho;\Fo) = \int^{\infty}_{\Fo} p_1(\rho,\F)\,d\F;
\end{equation}
this integral cannot be evaluated in closed form in terms of known special functions, but it is clear that the higher the optimal S/N, the higher the detection probability.

\subsection{False-alarm and detection probabilities for GW searches}
\label{sec:false}

In actual GW searches, the detector output will be correlated to a bank of optimal filters corresponding to different values of the intrinsic parameters $\xi^\mu$.
For a given set of detector data, the statistic $\F(\xi^\mu)$ is a generalized multiparameter random process known as \emph{random field} (see Adler's monograph \cite{A81} for a comprehensive discussion): we can use the theory of random fields to get a handle on the total false-alarm and detection probabilities for the entire filter bank.

We define the \emph{autocovariance} $\mathcal{C}$ of the random field $\F(\xi^\mu)$ as
\begin{equation}
\label{eq:C}
\mathcal{C}(\xi^\mu,{\xi'}^\mu) = {\cal E}_0[{\mathcal F(\xi^\mu)}{\mathcal F({\xi'}^\mu)}] - 
{\cal E}_0[{\mathcal F(\xi^\mu)}]{\cal E}_0[{\mathcal F({\xi'}^\mu)}],
\end{equation}
where the expectation value ${\cal E}_0$ is computed over an ensemble of realizations of noise (in absence of the signal). In Ref.\ \cite{JKS98} the total false-alarm probability was estimated by noticing that the autocovariance function tends to zero as the displacement $\Delta \xi^\mu = {\xi'}^\mu - \xi^\mu$ increases (and in fact, it is maximum for $\Delta \xi^\mu = 0$). The space of intrinsic parameters may then be partitioned into a set of \emph{elementary cells}, whereby the autocovariance is appreciably different from zero for within each cell, but negligible between cells. The number of elementary cells needed to cover the parameter space gives an estimate of the number of independent realizations of the random field (i.e., the number of statically independent ways that pure noise can be strongly correlated with one or more of the optimal filters).

There is of course some arbitrariness in choosing the boundary of the elementary cells; we define them by requiring that the autocovariance between the center and the surface be one half of the autocovariance \emph{at the center}:
\begin{equation}
\mathcal{C}(\xi^\mu,{\xi'}^\mu) = {\textstyle\frac{1}{2}}\mathcal{C}(\xi^\mu,\xi^\mu) \quad \mbox{for $\xi^\mu$ at cell center, ${\xi'}^\mu$ on cell boundary}.
\end{equation}
Taylor-expanding the autocovariance to second order in $\Delta \xi^\mu$, we obtain the approximate condition
\begin{equation}
\mathcal{C}(\xi^\mu,{\xi'}^\mu) \simeq
\mathcal{C}(\xi^\mu,\xi^\mu) + \left.\frac{1}{2}\frac{\partial^2 \mathcal{C}(\xi^\mu,{\xi'}^\mu)}
{\partial {\xi'}^\rho \partial {\xi'}^\sigma}\right|_{{\xi'}^\mu = {\xi}^\mu} \Delta {\xi}^\rho \Delta {\xi}^\sigma = {\textstyle\frac{1}{2}}\mathcal{C}(\xi^\mu,\xi^\mu),
\end{equation}
with implicit summation over $\rho$ and $\sigma$.
Within the approximation (necessary to obtain results in simple analytical form), the cell boundary is the (hyper-)ellipse defined by $G_{\rho \sigma} \Delta \xi^\rho \Delta \xi^\sigma = 1/2$, where \cite{O96}
\begin{equation}
\label{eq:G}
G_{\rho \sigma} = -\left.\frac{1}{2}\frac{1}{\mathcal{C}(\xi^\mu,\xi^\mu)}\frac{\partial^2 \mathcal{C}(\xi^\mu,{\xi'}^\mu)}
{\partial {\xi'}^\rho \partial {\xi'}^\sigma} \right|_{{{\xi'}^\mu} = \xi^\mu}
\end{equation}
(in Appendix \ref{app:r} we shall derive a relation between this $G_{\rho \sigma}$ and the Fisher information matrix).
The volume $V_{\text{cell}}$ of the elementary cell is then
\begin{equation}
\label{eq:vc}
V_{\text{cell}} = \frac{(\pi/2)^{K/2}}{\Gamma(K/2+1)\sqrt{\det G_{\rho \sigma}}},
\end{equation}
where $K$ is the number of intrinsic parameters, and $\Gamma$ is the gamma function.
The total number of elementary cells within the parameter volume $V$ is given by
\begin{equation}
N_\mathrm{cell} = \frac{\Gamma(K/2+1)}{(\pi/2)^{K/2}}\int_V\sqrt{\det G_{\rho \sigma}}\,dV.
\end{equation}
As discussed above, we consider the values of the statistic $\mathcal{F}$ within each cell as independent random variables, which in the absence of signal are distributed according to Eq.\ \eqref{p0}. By our definition of false alarms, the probability that $\F$ will not exceed the threshold $\F$ in a given cell is just $1 - P_F(\Fo)$; the probability that $\F$ will not exceed the threshold $\Fo$ \emph{in any of the cells} is
\begin{equation}
\label{FP}
1 - P_{F,\mathrm{tot}}(\Fo) \simeq [1 - P_F(\Fo)]^{N_\mathrm{cell}};
\end{equation}
this $P_{F,\mathrm{tot}}(\Fo)$ is therefore the total false-alarm probability for our detection scheme.

When the signal is present, a precise calculation of the probability
distribution function of ${\mathcal F}$ is nontrivial, since the
presence of the signal makes the random process ${\bf x}(t)$ nonstationary.
However, we can still use the detection probability given by Eq.\ \eqref{PD} for known intrinsic parameters as a substitute for the detection probability when the parameters are unknown. This is correct if we assume that, when the signal is present, the true values of the intrinsic parameters fall within the cell where ${\mathcal F}$ is maximum.  This approximation is accurate for sufficiently large S/N.

\section{Fast computation of the $\mathcal{F}$ statistic}
\label{ssec:alg}

The detection statistic $\F$ [Eq.\ \eqref{eq:mld3}] involves integrals of the general form
\begin{equation}
\int^{T_0}_0 x(t) \, m(t;\omega,\beta,\lambda) \,
\exp [i \phi_\mathrm{mod} (t;\omega,\dot{\omega},\beta,\lambda)] 
\exp [i \omega t] \, dt
\label{fstat}
\end{equation}
where $m$ is a combination of the complex modulation functions defined in App.\ A, 
while the phase modulation $\phi_\mathrm{mod}$ is given by
\begin{equation}
\label{eqs:MNA}
\phi_{mod}(t) = \frac{1}{2}\dot{\omega} t^2 + \omega R\cos\beta\cos(\Omega t + \eta_0 - \lambda)
\end{equation}
[Eq.\ \eqref{fase}]. We see that the integral \eqref{fstat} can be interpreted as a Fourier transform [and computed efficiently with a fast Fourier transform (FFT)], if $\phi_{mod}$ and $m$ do not depend on the frequency $\omega$.
In fact, even in that case we can still use FFTs by means of the procedure that we now present.

From the original data we generate several band-passed data sets, 
choosing the bandwidth of each set so that $m \exp [i\phi_{mod}]$ is approximately constant over the band. We then search for GW signals in each band-passed data set: this is done by computing the $\mathcal{F}$ statistic over a grid in the parameter space $(\dot{\omega}, \beta, \lambda)$, set finely enough that we do not miss any signal. We follow the grid-construction procedure presented in Sec.\ III A of Ref.\ \cite{ABJK02}.
The phase modulation can be usefully reparametrized as
\begin{equation}
\phi_{mod} = p_1 t^2  + A \cos(\Omega t) + B \sin(\Omega t), 
\end{equation}
where
\begin{equation}
\left\{
\begin{aligned}
p_1 &= {\textstyle \frac{1}{2}} \dot{\omega}, \\
A   &= \omega R \cos\beta\cos(\lambda - \eta_0), \\ 
B   &= \omega R \cos\beta\sin(\lambda - \eta_0).
\end{aligned}
\right.
\end{equation}
Since $m$ is a slowly changing function of time, we consider it constant
for the purpose of constructing a grid over the parameter space. 
The result is a uniform grid of prisms with hexagonal bases, where the parameter subspace $A$--$B$ is tiled by regular hexagons.
The grid in the parameters $\dot{\omega}$, $\beta$ and $\gamma$ is then derived by applying the inverse transformation,
\begin{equation}
\left\{
\begin{aligned}
\dot{\omega} &= 2 p_1, \\
\beta   &= \pm \arccos\left(\sqrt{A^2 + B^2}/\omega R\right), \\
\lambda &= \eta_0 + \arctan (B/A),
\end{aligned}
\right.
\label{eq:inverse}
\end{equation}
where for each band-passed data set we set the unknown frequency $\omega$ to the maximum frequency of the band. The computation of the ${\mathcal F}$ statistic includes both phase- and amplitude-modulation effects, even if these were neglected in the construction of the grid [in fact, the sign degeneracy for $\beta$ in Eq.\ \eqref{eq:inverse} is resolved by amplitude modulation, which distinguishes between sources in opposite directions with respect to the plane of the ecliptic].

Once we have a detection, the accurate estimation of signal parameters requires a second step. Since the coarse signal search described above is performed by evaluating the function $m \exp [i \phi_{mod}]$ at the maximum frequency of each band, our filters are not perfectly matched to the signal, and thus are not optimal; as a consequence, the location of the maximum of $\mathcal{F}$ does not correspond to the correct ML estimators. We therefore refine the coarse search by maximizing $\mathcal{F}$ near the coarse-search maximum, this time without any approximation.

\begin{figure}
\includegraphics[width=13cm]{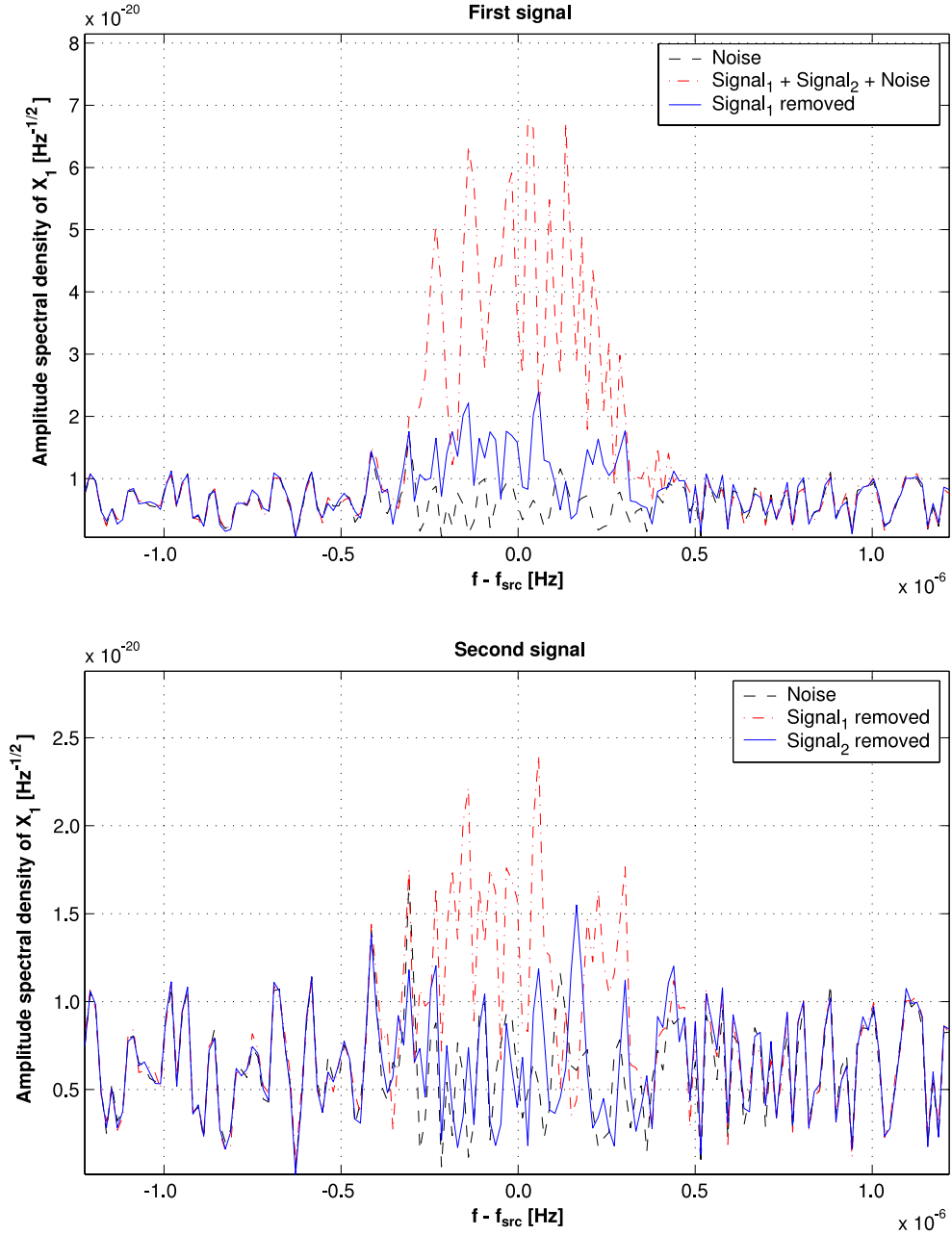}
\caption{Quality of signal reconstruction, as seen in the Fourier domain, in the first simulation. The original $X_1$ time series contains noise, pluse two monochromatic signals of equal frequency (3 mHz) and ecliptic longitude, but opposite ecliptic latitudes. We show the $X_1$ amplitude spectrum before and after the subtraction of the reconstructed signals, compared with the amplitude spectrum of noise alone. \label{sim1a}}
\end{figure}

\begin{figure}
\includegraphics[width=13cm]{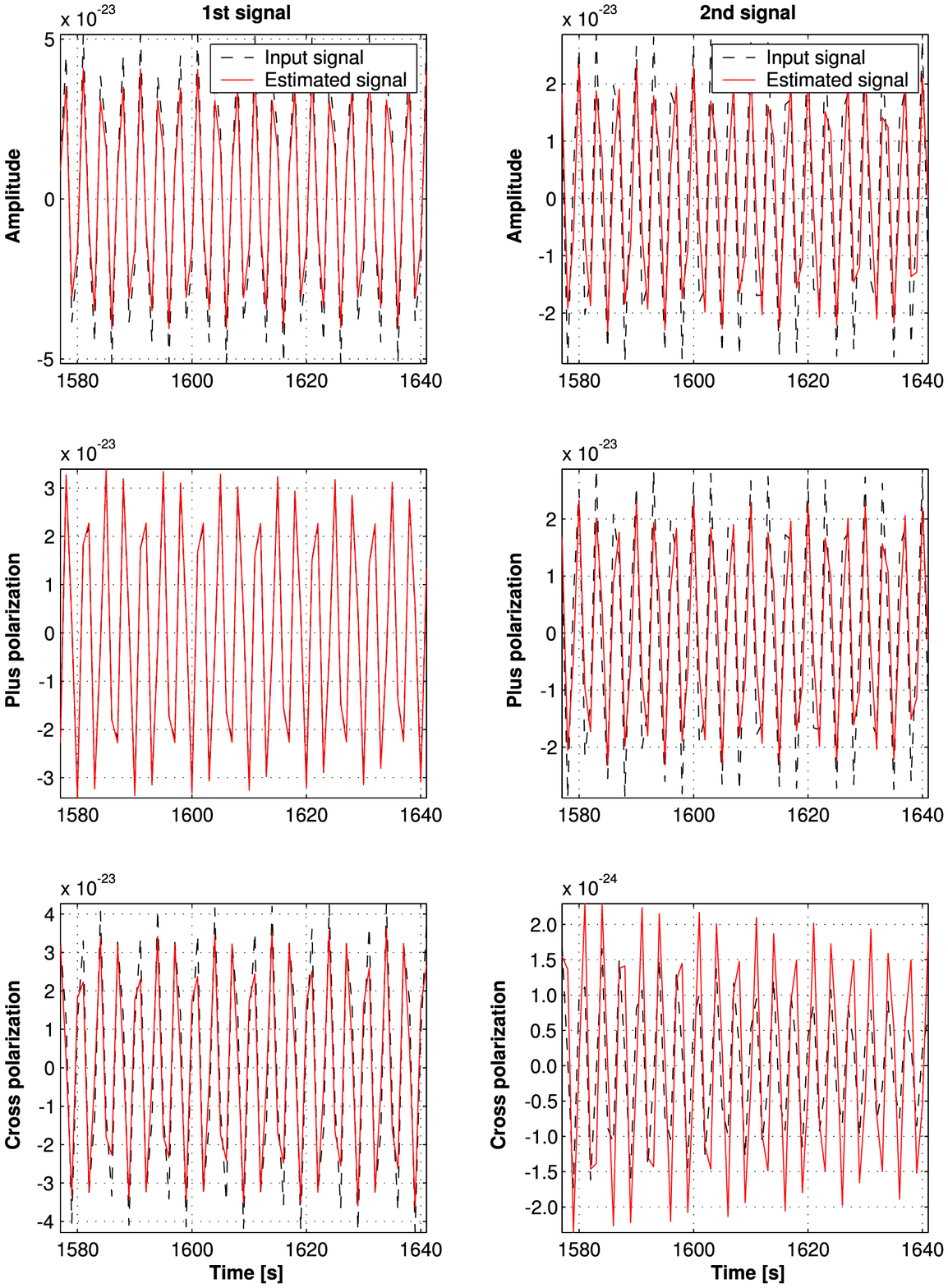}
\caption{Quality of signal reconstruction, as seen in the time domain, in the first simulation. The panels show the input signals (the stronger on the left, the weaker on the right), compared with the reconstructed signals; the two GW polarizations are plotted together (top row), and separately (middle and bottom rows).\label{sim1b}}
\end{figure}

We have performed a few numerical simulations to assess the performance of our optimal-filtering algorithm. Here we report on three of them.
In the first simulation we analyzed the $X_1$ TDI data corresponding to two simultaneous monochromatic signals, of frequency $f = 3$ mHz and S/Ns of 24 and 10, emitted from sources at opposite positions with respect to the plane of the ecliptic. We generated a one-year-long time series for $X_1$ by implementing Eq.\ \eqref{eq:x1lw} numerically, and we included noise by adding a Gaussian random process (as realized by a random number generator) with spectral density given by Eq.\ \eqref{eq:spec}.
We narrowbanded the $X_1$ data to a bandwidth of 0.125 mHz around 3 mHz, and we analyzed the resulting data by implementing the two-step procedure described above, using the Nelder-Mead maximization algorithm \cite{LRWW98} for the second step. The angular grid for the all-sky search consisted of about 900 points. We then performed the following operations: i) detecting the stronger signal and estimating its parameters; ii) reconstructing the stronger signal and subtracting it from the data; iii) detecting the weaker signal and estimating its parameters; iv) subtracting it from the data. Figure \ref{sim1a} shows the amplitude spectrum of $X_1$ before and after the subtraction of the two signals, as compared with the spectrum of noise alone. Figure \ref{sim1b} shows a comparison of the input signals with the reconstructed signals (built with the parameters specified by the ML estimators).  We see that the amplitude modulations in the GW response enable us to determine the sky location of two sources of the same frequency, and also to resolve the two GW polarizations. Signal resolution will degenerate rapidly as more sources of the same frequency are added, so the steps described above cannot be used as a general signal subtraction procedure \cite{notesub}.

\begin{figure}
  \includegraphics[width=11.5cm]{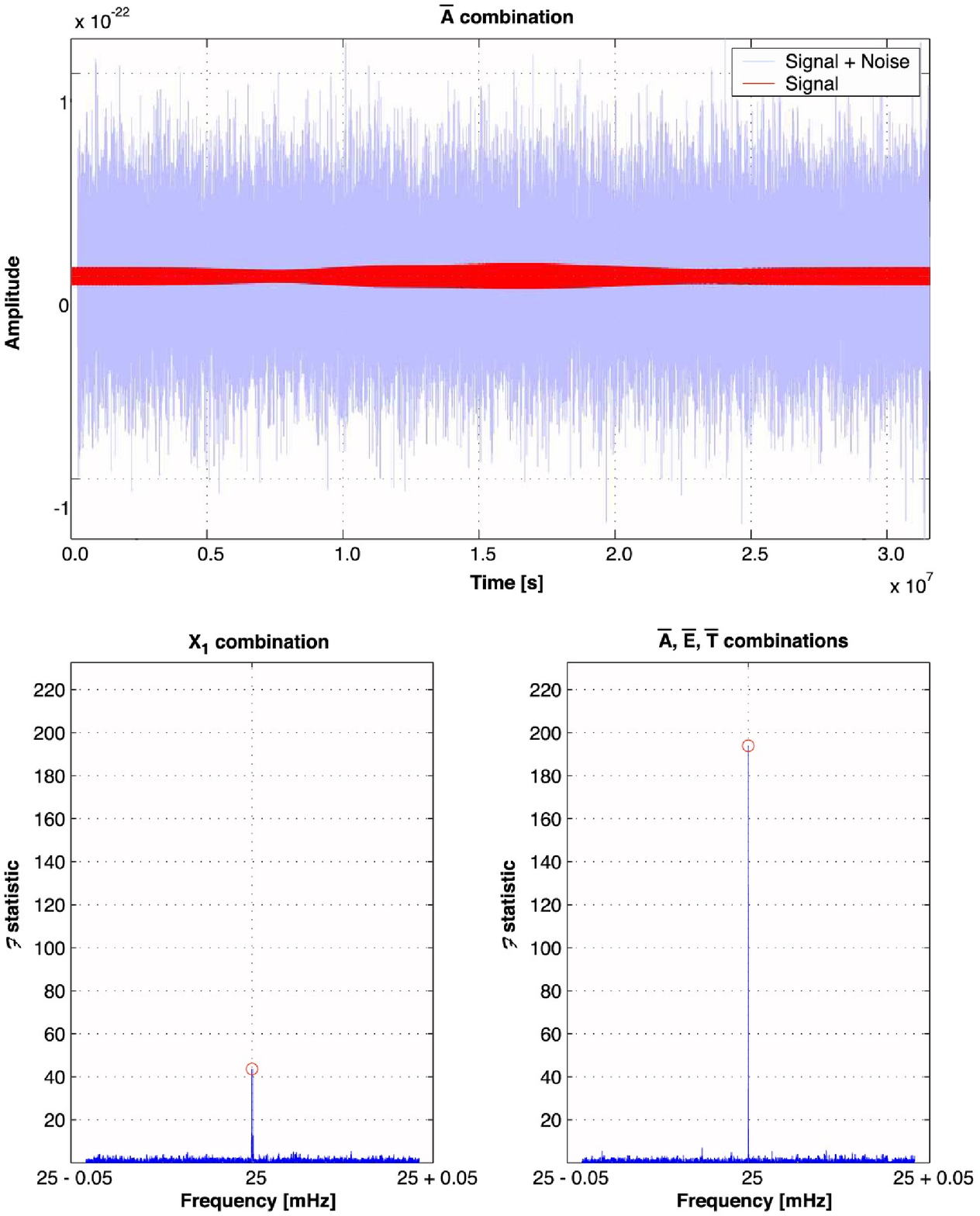}
  \caption{Maximum-likelihood detection in the second and third simulations. In the top panel we plot $\bar{A}(t)$ for the input signal alone, superimposed on the same observable for signal plus noise
In the bottom panel we plot the $\mathcal{F}$ statistic (already maximized over $\dot{f}$, $\beta$, and $\lambda$) near the input signal frequency, for a single-observable search using $X_1$, and for a multiple-observable search using $\bar{A}$, $\bar{E}$, $\bar{T}$. The frequency of the input signals is correctly estimated in both cases, but the statistical significance of the multiple-observable detection is higher.\label{sim3}}
\end{figure}

\begin{figure}
\includegraphics[width=13cm]{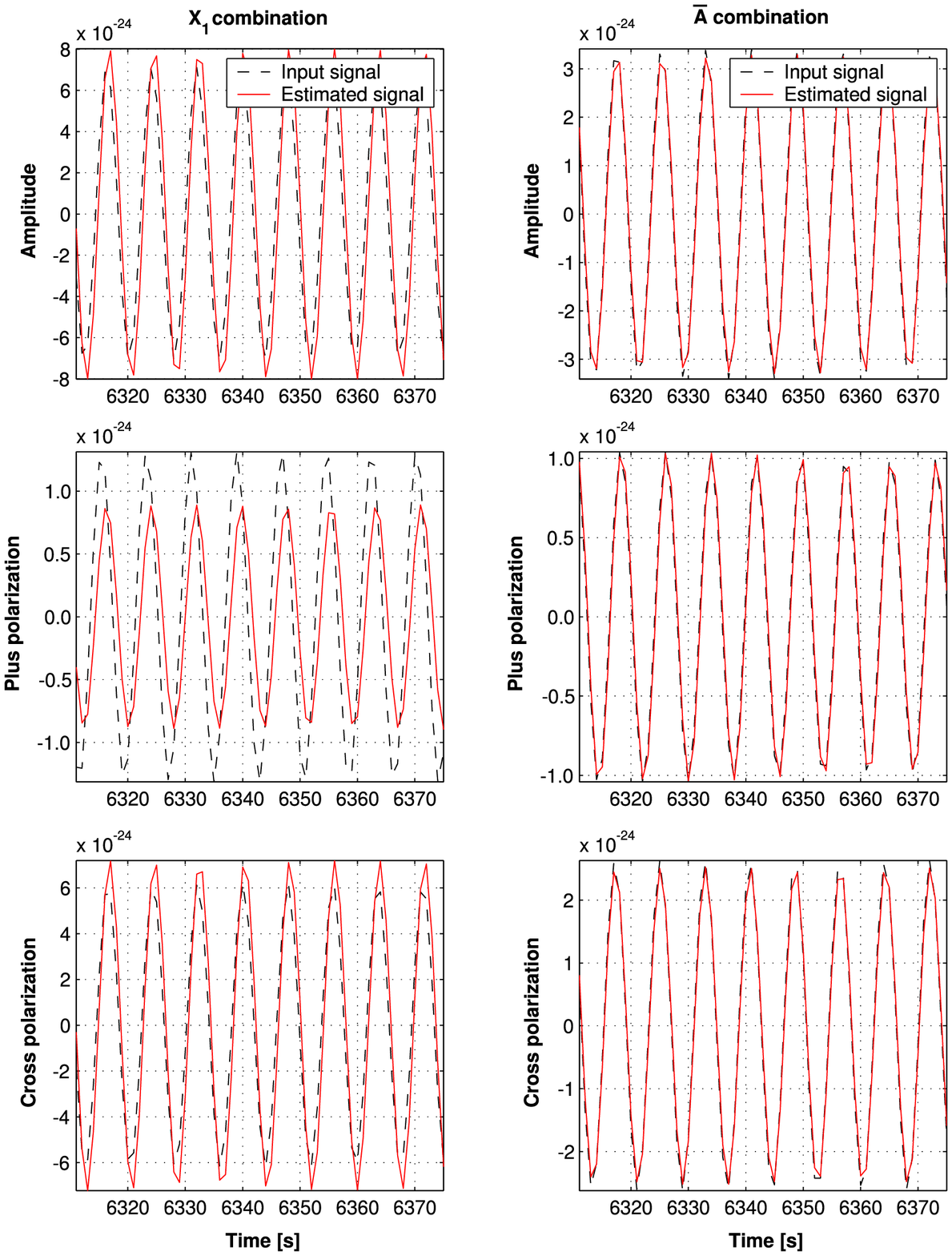}
\caption{Quality of signal reconstruction, as seen in the time domain, in the second and third simulations. The panels show the input signals ($X_1$ on the left, $\bar{A}$ on the right), compared with the reconstructed signals; the two GW polarizations are plotted together (top row), and separately (middle and bottom rows).
Signal reconstruction is more successful for the multiple-observable search (right) than for the single-observable search (left).
\label{sim2}}
\end{figure}

In the second simulation we analyzed the $X_1$ TDI data corresponding to a single signal of frequency $f = 25$mHz,  S/N = 9.5, and $\dot{f} = 6.5 \times 10^{-13}$ Hz s$^{-1}$ (corresponding to a binary of chirp mass $\mathcal{M}_c = 0.9 M_\odot$). We generated a one-year-long time series for $X_1$ by implementing numerically the exact GW response, Eq.\ \eqref{X_1}, and we added noise as described above.
We narrowbanded the $X_1$ data to a bandwidth of $0.5$ mHz, and again we analyzed the resulting data with the two-step procedure described above.
The sky search was performed on a small grid ($\sim 300$ gridpoints) around the true values of the signal parameters.
In the third simulation we analyzed the $\bar{A}$, $\bar{E}$, and $\bar{T}$ TDI data corresponding to the same signal, for a total S/N = 19. The ML search procedure was performed as in the second simulation.
The top panel of Fig.\ \ref{sim3} shows the TDI observable $\bar{A}(t)$ for the signal alone, superimposed on the TDI observable for signal plus noise: we see that the signal is more than one order of magnitude weaker than the noise. The bottom panel of Fig.\ \ref{sim3} shows the $\mathcal{F}$ statistic (already maximized over $\dot{f}$, $\beta$, and $\lambda$) near the input signal frequency. We see that the statistical significance is higher for the multiple-observable search than for the $X_1$ search. Figure \ref{sim2} shows a comparison of the input signals with the reconstructed signals.  We see that reconstruction is more accurate for the multiple-observable search, but in both searches our procedure resolves the two GW polarizations successfully.

We conclude that our proposed algorithm performs satifactorily, detecting the simulated signals, accurately estimating their parameters, and resolving the two GW polarizations, both in the low-frequency regime (first simulation) and high-frequency regime (second and third simulation). In a future paper, we plan to discuss in detail the expected errors in parameter estimation for a source with given frequency, sky position, and S/N.

\section*{Acknowledgments}
A.K.\ acknowledges support from the National Research Council under the Resident Research Associateship program at the
Jet Propulsion Laboratory, Caltech. M.V.'s research was supported by the LISA Mission Science Office at the Jet Propulsion Laboratory, Caltech. This research was performed at the Jet Propulsion Laboratory, California Institute of Technology, under contract with the National Aeronautics and Space Administration.

\appendix

\section{Complex representation of the response}
\label{app:c}

From Eqs.\ \eqref{alpha11} and \eqref{alpha14} it is easy to see that the Sagnac TDI observables $\alpha_i$ can be rewritten in the complex form
\begin{equation}
\alpha^\mathrm{GW}_i = 2 x \sin ({\textstyle \frac{3}{2}} x) \, \mathrm{Re} \left[
a^{(u)*} m^{(u)}_{\alpha_i}\!(t) \, e^{i \phi(t)} +
a^{(v)*} m^{(v)}_{\alpha_i}\!(t) \, e^{i \phi(t)}
\right],
\label{eq:complex}
\end{equation}
where $x = \omega L$, the complex amplitudes $a^{(u)}$ and $a^{(v)}$ are defined in Eq.\ \eqref{eq:CA} and \eqref{eq:CB}, the phase $\phi(t)$ by Eq.\ \eqref{fase}, and the complex \emph{modulation functions} $m^{(u)}_{\alpha_i}$ and $m^{(v)}_{\alpha_i}$ are given by
\begin{equation}
\label{eq:muv}
\left[\begin{array}{c}
m^{(u)}_{\alpha_i}(t) \\ m^{(v)}_{\alpha_i}(t)
\end{array}\right] =
\sum_{j=1}^3
\left[\begin{array}{c}
u_j(t) \\ v_j(t)
\end{array}\right] e^{-i x d_j} \left\{
\mathrm{sinc}\bigl[ (1+c_j)x/2 \bigr] n_{\alpha_i}^{+j} +
\mathrm{sinc}\bigl[ (1-c_j)x/2 \bigr] n_{\alpha_i}^{-j}
\right\},
\end{equation}
with $u_j(t)$, $v_j(t)$ given by Eqs.\ \eqref{ak} and \eqref{bk},
$c_j(t)$ and $d_j(t)$ by Eqs.\ \eqref{ci} and \eqref{di},
and with the constants $n_{\alpha_i}^{\pm j}$ given in the left part of Table \ref{tab:modfuncs}.
\begin{table*}
\begin{tabular}{c||c|c||c|c||c|c||c|c||c|c||c|c}
$j$ & $n_{\alpha_1}^{+j}$ & $n_{\alpha_1}^{-j}$
    & $n_{\alpha_2}^{+j}$ & $n_{\alpha_2}^{-j}$
    & $n_{\alpha_3}^{+j}$ & $n_{\alpha_3}^{-j}$
    & $n_{X_1}^{+j}$ & $n_{X_1}^{-j}$
    & $n_{X_2}^{+j}$ & $n_{X_2}^{-j}$
    & $n_{X_3}^{+j}$ & $n_{X_3}^{-j}$ \\
  \hline \hline
1 & $e^{-i 3 x}$ & $-e^{-i 3 x}$ & $e^{-i 2 x}$ & $-e^{-i 4 x}$ & $e^{-i 4 x}$ & $-e^{-i 2 x}$
& $0$ & $0$ & $e^{-i 7 x / 2}$ & $-e^{-i 9 x / 2}$ & $-e^{-i 9 x/2}$ & $-e^{-i 7 x /2}$ \\ 
2 & $e^{-i 2 x}$ & $-e^{-i 4 x}$ & $e^{-i 4 x}$ & $-e^{-i 2 x}$ & $e^{-i 3 x}$ & $-e^{-i 3 x}$
& $e^{-i 7 x / 2}$ & $e^{-i 9 x / 2}$ & $-e^{-i 9 x/2}$ & $-e^{-i 7 x/2}$ & $0$ & $0$ \\
3 & $e^{-i 4 x}$ & $-e^{-i 2 x}$ & $e^{-i 3 x}$ & $-e^{-i 3 x}$ & $e^{-i 2 x}$ & $-e^{-i 4 x}$
& $-e^{-i 9 x / 2}$ & $-e^{-i 7 x / 2}$ & $0$ & $0$ & $e^{-i 7 x/2}$ & $e^{-i 9 x/2}$
\end{tabular}
\caption{Constants that appear in the complex representation of the GW responses of the TDI observables. The constants $n_{\alpha_2}^{\pm j}$ and $n_{\alpha_3}^{\pm j}$ are obtained from $n_{\alpha_1}^{\pm j}$ by cyclical permutation of the index $j$, as are $n_{X_2}^{\pm j}$ and $n_{X_3}^{\pm j}$ from $n_{X_1}^{\pm j}$.\label{tab:modfuncs}}
\end{table*}

The optimal combinations $\bar{A}^\mathrm{GW}$, $\bar{E}^\mathrm{GW}$, $\bar{T}^\mathrm{GW}$ are given by formulas similar to Eq.\ \eqref{eq:complex}, with modulation functions
\begin{eqnarray}
m_{\bar{A}}^{(u)} &=& \frac{1}{\sqrt{2}} \left( m_{\alpha_3}^{(u)} - m_{\alpha_1}^{(u)} \right), \\
m_{\bar{E}}^{(u)} &=& \frac{1}{\sqrt{6}} \left( m_{\alpha_1}^{(u)} - 2 m_{\alpha_2}^{(u)} + m_{\alpha_3}^{(u)} \right), \\
m_{\bar{T}}^{(u)} &=& \frac{1}{\sqrt{3}} \left( m_{\alpha_1}^{(u)} + m_{\alpha_2}^{(u)} + m_{\alpha_3}^{(u)} \right),
\end{eqnarray}
and similar expressions for $m_{\bar{A}}^{(v)}$, $m_{\bar{E}}^{(v)}$, and $m_{\bar{T}}^{(v)}$.
The quantities $N^{(u)}$, $N^{(v)}$, $U$, $V$, and $W$ 
[Eqs.\ \eqref{eq:CFa}, \eqref{eq:CFb}, \eqref{eq:U}, \eqref{eq:V}, \eqref{eq:CQ}] that are needed to compute the ML amplitude estimators $\hat{a}^{(i)}$ and the $\mathcal{F}$ statistic [Eq. \eqref{eq:mld3}] can be written in terms of the complex modulation functions as
\begin{eqnarray} 
N^{(u)} &=& 4 x \sin ({\textstyle \frac{3}{2}} x)
\int_0^{T_0} \left[
\frac{\bar{A}(t)\, m_{\bar{A}}^{(u)}\!(t) + \bar{E}(t)\, m_{\bar{E}}^{(u)}\!(t)}{S_{\bar{A}}(\omega)} +
\frac{\bar{T}(t)\, m_{\bar{T}}^{(u)}\!(t)}{S_{\bar{T}}(\omega)}
\right] e^{i \phi(t)} dt, \\
N^{(v)} &=& 4 x \sin ({\textstyle \frac{3}{2}} x)
\int_0^{T_0} \left[
\frac{\bar{A}(t)\, m_{\bar{A}}^{(v)}\!(t) + \bar{E}(t)\, m_{\bar{E}}^{(v)}\!(t)}{S_{\bar{A}}(\omega)} +
\frac{\bar{T}(t)\, m_{\bar{T}}^{(v)}\!(t)}{S_{\bar{T}}(\omega)}
\right] e^{i \phi(t)} dt,
\end{eqnarray}
and
\begin{eqnarray}
U &=& 4 x^2 \sin^2 ({\textstyle \frac{3}{2}} x) \, (2/T_0)
\int_0^{T_0} \left[
\frac{|m_{\bar{A}}^{(u)}\!(t)|^2 + |m_{\bar{E}}^{(u)}\!(t)|^2}{S_{\bar{A}}(\omega)} +
\frac{|m_{\bar{T}}^{(u)}\!(t)|^2}{S_{\bar{T}}(\omega)}
\right] dt, \\
V &=& 4 x^2 \sin^2 ({\textstyle \frac{3}{2}} x) \, (2/T_0)
\int_0^{T_0} \left[
\frac{|m_{\bar{A}}^{(v)}\!(t)|^2 + |m_{\bar{E}}^{(v)}\!(t)|^2}{S_{\bar{A}}(\omega)} +
\frac{|m_{\bar{T}}^{(v)}\!(t)|^2}{S_{\bar{T}}(\omega)}
\right] dt, \\
W &=& 4 x^2 \sin^2 ({\textstyle \frac{3}{2}} x) \, (2/T_0)
\int_0^{T_0} \left[
\frac{m_{\bar{A}}^{(u)*}\!(t) m_{\bar{A}}^{(v)}\!(t) + m_{\bar{E}}^{(u)*}\!(t) m_{\bar{E}}^{(v)}\!(t)}{S_{\bar{A}}(\omega)} +
\frac{m_{\bar{T}}^{(u)*}\!(t) m_{\bar{T}}^{(v)}\!(t)}{S_{\bar{T}}(\omega)}
\right] dt.
\end{eqnarray}
The $X_i$ TDI observables can be written in the complex form as
\begin{equation}
X^\mathrm{GW}_i =
4 x \sin(x) \sin(2x) \, \mathrm{Re} \left[
i a^{(u)*} m^{(u)}_{X_i}\!(t) \, e^{i \phi(t)} +
i a^{(v)*} m^{(v)}_{X_i}\!(t) \, e^{i \phi(t)}
\right];
\label{eq:xcomplex}
\end{equation}
the modulation functions $m^{(u)}_{X_i}\!(t)$ and $m^{(v)}_{X_i}\!(t)$ have exactly the same functional form as the functions $m^{(u)}_{\alpha_i}\!(t)$, $m^{(v)}_{\alpha_i}\!(t)$ defined in Eq.\ \eqref{eq:muv}, except that the coefficients $n_{X_i}^{\pm j}$ are those given in the right part of Table \ref{tab:modfuncs}. \emph{For the single $X_1$ observable}, the ML estimators for the amplitudes and for $\mathcal{F}$ are again given by
\begin{eqnarray}
\label{eq:mld1x}
\hat{a}^{(u)} &=&  2 \, (T_0 \Delta_{X_1})^{-1} \bigl[V_{X_1} N^{(u)}_{X_1} - W^*_{X_1}  N_{X_1}^{(v)} \bigr], \\
\label{eq:mld2x}
\hat{a}^{(v)} &=&  2 \, (T_0 \Delta_{X_1})^{-1} \bigl[U_{X_1} N^{(v)}_{X_1} - W_{X_1} \, N_{X_1}^{(u)} \bigr],
\end{eqnarray}
[where $\Delta_{X_1} = U_{X_1} V_{X_1} - |W_{X_1}|^2$] and
\begin{equation}
\label{eq:mld3x}
{\mathcal F} = (T_0 \Delta)^{-1} \left\{ V_{X_1} \bigl|N_{X_1}^{(u)}\bigr|^2 + U_{X_1} \bigl|N_{X_1}^{(v)}\bigr|^2 - 2 \, \mathrm{Re} \left[ W_{X_1} \, N_{X_1}^{(u)} (N_{X_1}^{(v)})^* \right] \right\},
\end{equation}
with
\begin{eqnarray}
N^{(u)}_{X_1} &=& 8 x \sin(x) \sin(2x)
\int_0^{T_0} \left[
\frac{X_1(t) \, m_{X_1}^{(u)}\!(t)}{S_{X_1}\!(\omega)}
\right] i e^{i \phi(t)} dt, \\
N^{(v)}_{X_1} &=& 8 x \sin(x) \sin(2x)
\int_0^{T_0} \left[
\frac{X_1(t) \, m_{X_1}^{(v)}\!(t)}{S_{X_1}\!(\omega)}
\right] i e^{i \phi(t)} dt,
\end{eqnarray}
and
\begin{eqnarray}
U_{X_1} &=& 16 x^2 \sin^2(x) \sin^2(2x) \, (2/T_0)
\int_0^{T_0} \left[
\frac{|m_{X_1}^{(u)}\!(t)|^2}{S_{X_1}\!(\omega)}
\right] dt, \\
V_{X_1} &=& 16 x^2 \sin^2(x) \sin^2(2x) \, (2/T_0)
\int_0^{T_0} \left[
\frac{|m_{X_1}^{(v)}\!(t)|^2}{S_{X_1}\!(\omega)}
\right] dt, \\
W_{X_1} &=& 16 x^2 \sin^2(x) \sin^2(2x) \, (2/T_0)
\int_0^{T_0} \left[
\frac{m_{X_1}^{(u)*}\!(t) m_{X_1}^{(v)}\!(t)}{S_{X_1}\!(\omega)}
\right] dt.
\end{eqnarray}

\section{Reduced information matrix}
\label{app:r}

It is interesting to examine the relation between the matrix $G^{\mu \nu}$ defined by Eq.\ \eqref{eq:G} and the Fisher information matrix $\Gamma^{ij}$. We consider the case of a single TDI observable; multiple observables can be treated in similar fashion. As seen in Sec.\ II, the generic TDI GW response $h(t)$ can be written as the linear combination
\begin{equation}
h(t) = \sum^4_{k=1} a^{(k)} h^{(k)}(t,\xi^\mu);
\label{eq:siga}
\end{equation}
as discussed in Sec.\ \ref{sec:ml}, the amplitudes $a^{(k)}$ are extrinsic parameters, while all the other parameters (denoted together as $\xi^\mu$) are intrinsic [all the parameters are denoted together as $\theta^i \equiv (a^{(k)}, \xi^\mu)$]. Note that in the case of the TDI observables $X_1^\mathrm{GW}$ or $\bar{A}^\mathrm{GW}$ [Eqs.\ \eqref{X_1}, \eqref{eq:aetdecomp}], the component functions $h^{(k)}(t)$ would include the factors $4 x \sin(x) \sin(2x)$ and $2 x \sin(\frac{3}{2}x)$ respectively.

In this notation, it is easy to show that the optimal S/N and the Fisher matrix can be written as
\begin{equation}
\label{eq:apprho2}
\rho^2 = \mathbf{a}^T \cdot {\sf M} \cdot \mathbf{a},
\end{equation}
and
\begin{eqnarray}
\Gamma^{ij} =
\left(\begin{array}{cc}
{\sf M} & {\sf F} \cdot \mathbf{a} \\
\mathbf{a}^T \cdot {\sf F}^T  & \mathbf{a}^T \cdot {\sf S} \cdot \mathbf{a}
\end{array}\right),
\label{G}
\end{eqnarray} 
where the top and left blocks correspond to the extrinsic parameters, while the bottom and right blocks correspond to the intrinsic parameters.
The superscript ``${}^T$'' denotes transposition over the extrinsic parameter indices.
Furthermore, $\mathbf{a} \equiv (a^{(1)}, a^{(2)}, a^{(3)}, a^{(4)})$, and the matrices ${\sf M}$, ${\sf F}$, and ${\sf S}$ are given by
\begin{gather}
M^{(k)(l)} = (h^{(k)}|h^{(l)}), \\
F^{(k)(l)}_\mu = \bigg(h^{(k)}\bigg|\frac{\partial h^{(l)}}{\partial \xi^\mu}\bigg), \\
S^{(k)(l)}_{\mu\nu} = \bigg(\frac{\partial h^{(k)}}{\partial \xi^\mu}\bigg|\frac{\partial h^{(l)}}{\partial \xi^\nu}\bigg).
\end{gather}
The covariance matrix $C^{ij}$, which expresses the expected variance of the ML parameter estimators, is defined as $(\Gamma^{-1})^{ij}$. Using the standard formula for the inverse of a block matrix \cite{Meyer} we have
\bea
{\sf C} =
\left(\begin{array}{cc}
{\sf M}^{-1} + {\sf M}^{-1} \cdot  ({\sf F} \cdot \mathbf{a}) \cdot \bar{\Gamma}^{-1} \cdot ({\sf F} \cdot \mathbf{a})^T \cdot {\sf M}^{-1}  &  
- {\sf M}^{-1} \cdot ({\sf F} \cdot \mathbf{a}) \cdot \bar{\Gamma}^{-1} \\
- \bar{\Gamma}^{-1} \cdot ({\sf F} \cdot \mathbf{a})^T \cdot {\sf M}^{-1}  &  \bar{\Gamma}^{-1}
\end{array}\right),
\label{eq:CC}
\eea 
where
\be
\bar{\Gamma} = \mathbf{a}^T \cdot ({\sf S} - {\sf F}^T \cdot {\sf M}^{-1} \cdot {\sf F}) \cdot \mathbf{a}.
\ee 
We shall call $\bar{\Gamma}^{\mu \nu}$ (the \emph{Schur complement} of ${\sf M}$) the \emph{projected Fisher matrix} (onto the space of intrinsic parameters).
Because the projected Fisher matrix is the inverse of the intrinsic-parameter submatrix of the covariance matrix $C^{ij}$, it expresses the information available about the intrinsic parameters once the extrinsic parameters are set to their ML estimators. Note that $\bar{\Gamma}^{\mu \nu}$ is still a function of the putative extrinsic parameters.
Using Eq.\ \eqref{eq:apprho2} we define the \emph{normalized projected Fisher matrix}
\begin{equation}
\bar{\Gamma}_n \equiv \bar{\Gamma} / \rho^2 = \frac{\mathbf{a}^T \cdot ({\sf S} - {\sf F}^T \cdot {\sf M}^{-1} \cdot {\sf F}) \cdot \mathbf{a}}{\mathbf{a}^T \cdot {\sf M} \cdot \mathbf{a}}.
\end{equation}
From the Rayleigh principle \cite{Meyer}, it follows that the minimum value of
the component $\bar{\Gamma}_n^{\mu\nu}$ is given by the smallest 
eigenvalue (taken with respect to the extrinsic parameters) of the matrix $[({\sf S} - {\sf F}^T \cdot {\sf M}^{-1} \cdot {\sf F}) \cdot {\sf M}^{-1}]^{\mu \nu}$. Similarly, the maximum value of the component $\bar{\Gamma}_n^{\mu\nu}$ is given by the largest eigenvalue of that matrix. Because the trace of a matrix is equal to the sum of its eigenvalues, the matrix
\begin{equation}
\tilde{\Gamma} = \frac{1}{4} \mathrm{Tr} \, [({\sf S} - {\sf F}^T \cdot {\sf M}^{-1} \cdot {\sf F}) \cdot {\sf M}^{-1}],
\end{equation}
where the trace is taken over the extrinsic-parameter indices, expresses the information available about the intrinsic parameters, averaged over the possible values of the extrinsic parameters. Note that the factor $\frac{1}{4}$ is specific to the case of four extrinsic parameters. We shall call $\tilde{\Gamma}^{\mu \nu}$ the \emph{reduced Fisher matrix}. This matrix is a function of the intrinsic parameters alone.

Let us now compute the components of $G^{\mu \nu}$, defined by Eq.\ \eqref{eq:G}. 
We start from $\log \Lambda$, which in our notation is given by
\begin{equation}
\log\Lambda = \mathbf{a}^T \cdot {\sf N} - {\textstyle \frac{1}{2}} \, \mathbf{a}^T \cdot {\sf M} \cdot \mathbf{a},
\end{equation}
where $N^{(k)} = (x|h^{(k)})$, with $x(t) = n(t) + h(t)$, and $n(t)$ a zero-mean Gaussian random process. The ML estimators $\hat{a}^{(k)}$ are given by 
$\hat{\mathbf{a}} = {\sf M}^{-1} \cdot {\sf N}$,
so for the $\F$ statistic we have
$\F = {\textstyle \frac{1}{2}} \, {\sf N}^T \cdot {\sf M}^{-1} \cdot {\sf N}$.
Using the relations \cite{JK00}
\begin{gather}
\label{for1a}
\mathcal{E}\left\{(n|s_1)(n|s_2)\right\} = (s_1|s_2), \\
\mathcal{E}\left\{(n|s_1)(n|s_2)(n|s_3)(n|s_4)\right\}
= (s_1|s_2)(s_3|s_4) + (s_1|s_3)(s_2|s_4) + (s_1|s_4)(s_2|s_3),
\end{gather}
where $s_1$, $s_2$, $s_3$, and $s_4$ are deterministic functions,
we find that the autocovariance function $\mathcal{C}(\xi^\mu,{\xi'}^\mu)$ of Eq.\ \eqref{eq:C} is given by
\begin{equation}
\mathcal{C}(\xi^\mu,{\xi'}^\mu) = \frac{1}{2} \mathrm{Tr} \, [{\sf Q}^T \cdot {\sf M}^{-1} \cdot {\sf Q} \cdot {\sf M}'^{-1}],
\label{eq:cqm}
\end{equation}
where
\begin{equation}
Q_{kl} = (h^{(k)}|h'^{(l)}),
\end{equation}
and the primes denote functions of the primed parameters ${\xi'}^\mu$. Inserting Eq.\ \eqref{eq:cqm} into Eq.\ \eqref{eq:G}, after some lengthy algebra (omitted here) we come to the final result
\begin{equation}
G_{\mu \nu} = \frac{1}{4} \sum_{k,l,m,n} 
\bigl[S^{(k)(l)}_{\mu \nu} - F^{(m)(k)}_{\mu} (M^{-1})^{(m)(n)} F^{(n)(l)}_\nu\bigr] (M^{-1})^{(l)(k)}
= \tilde{\Gamma}_{\mu \nu}.
\end{equation}
Thus, the $\mathcal{F}$-statistic metric $G^{\mu \nu}$ \cite{O96} is found to be exactly equal to the reduced Fisher matrix $\tilde{\Gamma}^{\mu \nu}$; that this should be the case is understandable, since both matrices contain information about the relatedness of waveforms with nearby values of their intrinsic parameters (while both assume that the extrinsic parameters are being set to their ML estimators). For a related argument about the placement of templates for a partially maximized detection statistic, see Ref.\ \cite{pbcv}.

\section{First-generation TDI responses}
\label{app:f}

The GW response of the first-generation TDI observable $X$ is given by \cite{AET99}
\begin{eqnarray}
X^\mathrm{GW} & = &
  (y^\mathrm{GW}_{31} + y^\mathrm{GW}_{13,2}) 
+ (y^\mathrm{GW}_{21} + y^\mathrm{GW}_{12,3})_{,22} 
- (y^\mathrm{GW}_{21} + y^\mathrm{GW}_{12,3}) 
- (y^\mathrm{GW}_{31} +  y^\mathrm{GW}_{13,2})_{,33};
\end{eqnarray}
after some algebra we get to
\begin{equation}
X^\mathrm{GW} (t) = 2 \, \omega L \, \sin(\omega L) \sum_{k=1}^{4} a^{(k)} X^{(k)}(t),
\label{X}
\end{equation}
where the functions $X^{(k)}(t)$ are given by
\begin{eqnarray}
\left[\!\begin{array}{c}
X^{(1)} \\
X^{(2)}
\end{array}\!\right]
&=&
\left[\!\begin{array}{c}
u_2(t) \\
v_2(t)
\end{array}\!\right]\!
\bigl\{\mbox{sinc}\bigl[(1+c_2)x/2\bigr]\cos\bigl[\phi(t) - x d_2 - 3x/2\bigr]    
              + \mbox{sinc}\bigl[(1-c_2)x/2\bigr]\cos\bigl[\phi(t) - x d_2 -
              5x/2\bigr]\bigr\} \nonumber
\\ 
&& -  
\left[\!\begin{array}{c}
u_3(t) \\
v_3(t)
\end{array}\!\right]\!
\bigl\{\mbox{sinc}\bigl[(1+c_3)x/2\bigr]\cos\bigl[\phi(t) - x d_3 - 5x/2\bigr] 
              + \mbox{sinc}\bigl[(1-c_3)x/2\bigr]\cos\bigl[\phi(t) - x d_3 -
             3x/2\bigr]\bigr\}, \\ 
\left[\!\begin{array}{c}
X^{(3)} \\
X^{(4)}
\end{array}\!\right]
&=&
\left[\!\begin{array}{c}
u_2(t) \\
v_2(t)
\end{array}\!\right]\!
\bigl\{\mbox{sinc}\bigl[(1+c_2)x/2\bigr]\sin\bigl[\phi(t) - x d_2 - 3x/2\bigr]    
              + \mbox{sinc}\bigl[(1-c_2)x/2\bigr]\sin\bigl[\phi(t) - x d_2 - 5x/2\bigr]\bigr\} \nonumber \\ 
&& -
\left[\!\begin{array}{c}
u_3(t) \\
v_3(t)
\end{array}\!\right]\!
\bigl\{\mbox{sinc}\bigl[(1+c_3)x/2\bigr]\sin\bigl[\phi(t) - x d_3 - 5x/2\bigr] 
              + \mbox{sinc}\bigl[(1-c_3)x/2\bigr]\sin\bigl[\phi(t) - x d_3 -
             3x/2\bigr]\bigr\},
\label{X_last}
\end{eqnarray}
where $x = \omega L$ and ${\mathrm{sinc}\,\ldots = (\sin \ldots)/(\ldots)}$. 
The GW responses for $Y$ and $Z$ can be obtained by cyclical permutation of the spacecraft indices.

The GW response for $\alpha$ can be written in similar form:
\begin{equation}
\alpha^\mathrm{GW} (t) = \omega L \sum_{k=1}^{k=4} a^{(k)} \alpha^{(k)}(t),
\label{alpha}
\end{equation}
where
\begin{eqnarray}
\left[\!\begin{array}{c}
\alpha^{(1)} \\
\alpha^{(2)}
\end{array}\!\right]
&=&
\left[\!\begin{array}{c}
u_1(t) \\ v_1(t)
\end{array}\!\right]
\bigl\{\mbox{sinc}\bigl[(1+c_1)x/2\bigr]\sin\bigl[\phi(t) - x d_1 - 3x/2\bigr] 
- \mbox{sinc}\bigl[(1-c_1)x/2\bigr]\sin\bigl[\phi(t) - x d_1 - 3x/2\bigr]\bigr\} \nonumber \\ 
&& + 
\left[\!\begin{array}{c}
u_2(t) \\ v_2(t)
\end{array}\!\right]
\bigl\{\mbox{sinc}\bigl[(1+c_2)x/2\bigr]\sin\bigl[\phi(t) - x d_2 - x/2\bigr]   
-    \mbox{sinc}\bigl[(1-c_2)x/2\bigr]\sin\bigl[\phi(t) - x
d_2 - 5 x/2\bigr]\bigr\} \label{alpha1} \\ 
&& + 
\left[\!\begin{array}{c}
u_3(t) \\ v_3(t)
\end{array}\!\right]
\bigl\{\mbox{sinc}\bigl[(1+c_3)x/2\bigr]\sin\bigl[\phi(t) - x d_3 - 5 x/2\bigr] 
-    \mbox{sinc}\bigl[(1-c_3)x/2\bigr]\sin\bigl[\phi(t) - x
d_3 - x/2\bigr]\bigr\}, \nonumber \\ 
\left[\!\begin{array}{c}
\alpha^{(3)} \\ \alpha^{(4)}
\end{array}\!\right]
&=& -
\left[\!\begin{array}{c}
u_1(t) \\ v_1(t)
\end{array}\!\right]
\bigl\{\mbox{sinc}\bigl[(1+c_1)x/2\bigr]\cos\bigl[\phi(t) - x d_1 - 3 x/2\bigr]  
- \mbox{sinc}\bigl[(1-c_1)x/2\bigr]\cos\bigl[\phi(t) - x
d_1 - 3 x/2\bigr]\bigr\} \nonumber \\
&& - 
\left[\!\begin{array}{c}
u_2(t) \\ v_2(t)
\end{array}\!\right]
\bigl\{\mbox{sinc}\bigl[(1+c_2)x/2\bigr]\cos\bigl[\phi(t) - x d_2 - x/2\bigr]    
- \mbox{sinc}\bigl[(1-c_2)x/2\bigr]\cos\bigl[\phi(t) - x d_2 - 5 x/2\bigr]\bigr\} \label{alpha4} \\ 
&& -
\left[\!\begin{array}{c}
u_3(t) \\ v_3(t)
\end{array}\!\right]
\bigl\{\mbox{sinc}\bigl[(1+c_3)x/2\bigr]\cos\bigl[\phi(t) - x d_3 - 5 x/2\bigr] 
- \mbox{sinc}\bigl[(1-c_3)x/2\bigr]\cos\bigl[\phi(t) - x
d_3 - x/2\bigr]\bigr\}. \nonumber
\end{eqnarray}
The $\beta$ and $\gamma$ combinations are again obtained by cyclical permutation of the spacecraft indices.

For the TDI observable $\zeta$ we find 
\begin{equation}
\zeta^\mathrm{GW} (t) = \omega L \sum_{k=1}^{4} a^{(k)} \zeta^{(k)}(t),
\label{zeta}
\end{equation}
with
\begin{eqnarray}
\left[\!\begin{array}{c}
\zeta^{(1)} \\ \zeta^{(2)}
\end{array}\!\right]
&=&
\left[\!\begin{array}{c}
u_1(t) \\ v_1(t)
\end{array}\!\right]
\bigl\{\mbox{sinc}\bigl[(1+c_1)x/2\bigr]\sin\bigl[\phi(t) - x d_1 - 3 x/2\bigr] 
- \mbox{sinc}\bigl[(1-c_1)x/2\bigr]\sin\bigl[\phi(t) - x
d_1 - 3 x/2\bigr]\bigr\} \nonumber \\ 
&& + 
\left[\!\begin{array}{c}
u_2(t) \\ v_2(t)
\end{array}\!\right]
\bigl\{\mbox{sinc}\bigl[(1+c_2)x/2\bigr]\sin\bigl[\phi(t) - x d_2 - 3 x/2\bigr]   
-    \mbox{sinc}\bigl[(1-c_2)x/2\bigr]\sin\bigl[\phi(t) - x
d_2 - 3 x/2\bigr]\bigr\} \label{zeta1} \\ 
&& + 
\left[\!\begin{array}{c}
u_3(t) \\ v_3(t)
\end{array}\!\right]
\bigl\{\mbox{sinc}\bigl[(1+c_3)x/2\bigr]\sin\bigl[\phi(t) - x d_3 - 3 x/2\bigr] 
-    \mbox{sinc}\bigl[(1-c_3)x/2\bigr]\sin\bigl[\phi(t) - x
d_3 - 3 x/2\bigr]\bigr\}, \nonumber \\ 
\left[\!\begin{array}{c}
\zeta^{(3)} \\ \zeta^{(4)}
\end{array}\!\right]
&=& -
\left[\!\begin{array}{c}
u_1(t) \\ v_1(t)
\end{array}\!\right]
\bigl\{\mbox{sinc}\bigl[(1+c_1)x/2\bigr]\cos\bigl[\phi(t) - x d_1 - 3 x/2\bigr]  
- \mbox{sinc}\bigl[(1-c_1)x/2\bigr]\cos\bigl[\phi(t) - x
d_1 - 3 x/2\bigr]\bigr\} \nonumber \\
&& -
\left[\!\begin{array}{c}
u_2(t) \\ v_2(t)
\end{array}\!\right]
\bigl\{\mbox{sinc}\bigl[(1+c_2)x/2\bigr]\cos\bigl[\phi(t) - x d_2 - 3 x/2\bigr]    
- \mbox{sinc}\bigl[(1-c_2)x/2\bigr]\cos\bigl[\phi(t) - x 
d_2 - 3 x/2\bigr]\bigr\}  \label{zeta3} \\ 
&& -
\left[\!\begin{array}{c}
u_3(t) \\ v_3(t)
\end{array}\!\right]
\bigl\{\mbox{sinc}\bigl[(1+c_3)x/2\bigr]\cos\bigl[\phi(t) - x d_3 - 3 x/2\bigr] 
- \mbox{sinc}\bigl[(1-c_3)x/2\bigr]\cos\bigl[\phi(t) - x d_3 - 3 x/2\bigr]\bigr\}. \nonumber
\end{eqnarray}
Finally, the optimal TDI observables $A$, $E$, and $T$ \cite{PTLA02}
are defined as linear combinations of $\alpha$, $\beta$ and $\gamma$:
\begin{eqnarray}
A &=& \frac{1}{\sqrt{2}}(\gamma - \alpha), \nonumber \\
E &=& \frac{1}{\sqrt{6}}(\alpha - 2\beta + \gamma), \label{eq:aetfirst} \\
T &=& \frac{1}{\sqrt{3}}(\alpha +  \beta + \gamma). \nonumber
\end{eqnarray}

The long-wavelength approximation to the GW responses is obtained by taking 
the leading-order terms of the generic expressions 
in the limit of $\omega L \rightarrow 0$. For instance, for $X$ [Eqs.\ \eqref{X}--\eqref{X_last}], we get
\begin{equation}
X^\mathrm{GW}_{\mathrm{LW}} \simeq 4 (\omega L)^2 \left\{
\left[ u_2(t) - u_3(t) \right] \left[ a^{(1)} \cos \phi(t) + a^{(3)} \sin \phi(t) \right]+
\left[ v_2(t) - v_3(t) \right] \left[ a^{(2)} \cos \phi(t) + a^{(4)} \sin \phi(t) \right]
\right\},
\label{eq:xlw}
\end{equation}
with $a^{(k)}$ given by Eqs.\ \eqref{eq:ampone}--\eqref{Aamp}, and $u_i(t)$, $v_i(t)$ by Eqs.\ \eqref{ak}, \eqref{bk}. The LW responses for $Y$ and $Z$ can be obtained by cyclical permutation of the indices. 
Adopting the notation of Ref.\ \cite{AET99}, we find also that
\begin{equation}
X^\mathrm{GW}_{\mathrm{LW}} (t) \simeq 2 L^2 \left[(\hat{\bf n}^L_3)' \cdot \ddot{{\sf H}}^L(t) \cdot \hat{\bf n}^L_3 
- (\hat{\bf n}^L_2)' \cdot \ddot{{\sf H}}^L(t) \cdot \hat{\bf n}^L_2\right],
\end{equation}
where ``$\ddot{\phantom{x}}$'' denotes the second time derivative,
$\hat{\bf n}^L_i$ is given by Eq.\ \eqref{eq:nlisa},
and ${\sf H}^L(t)$ is given by Eq.\ \eqref{HT}.

The GW response of the Sagnac observable $\alpha^\mathrm{GW}_{\mathrm{LW}}$ 
is equal simply to $\frac{1}{2} X^\mathrm{GW}_{\mathrm{LW}}$. 
From Eqs.\ \eqref{eq:aetfirst} we then get the LW GW responses 
$A_\mathrm{LW}^\mathrm{GW}$, $E_\mathrm{LW}^\mathrm{GW}$, 
and $T_\mathrm{LW}^\mathrm{GW}$:
\begin{align}
A^\mathrm{GW}_{\mathrm{LW}} &\simeq \sqrt{2} \, (\omega L)^2 \left\{
\left[ - 2 u_2(t) + u_1(t) + u_3(t) \right] \left[ a^{(1)} \cos \phi(t) + a^{(3)} \sin \phi(t) \right] \right. \nonumber \\
& \phantom{\simeq \sqrt{2} \, (\omega L)^2 \left\{ \left[ - 2 u_2(t) + u_1(t) + u_3(t) \right] \left[ a^{(1)} \cos\right. \right.} + \left. \left[ - 2 v_2(t) + v_1(t) + v_3(t) \right] \left[ a^{(2)} \cos \phi(t) + a^{(4)} \sin \phi(t) \right] \right\}, \label{eq:mylwa1} \\
E^\mathrm{GW}_{\mathrm{LW}} &\simeq \sqrt{6} \, (\omega L)^2 \left\{
\left[ u_1(t) - u_3(t) \right] \left[ a^{(1)} \cos \phi(t) + a^{(3)} \sin \phi(t) \right] + \left[ v_1(t) - v_3(t) \right] \left[ a^{(2)} \cos \phi(t) + a^{(4)} \sin \phi(t) \right] \right\}, \label{eq:mylwe1} \\
T^\mathrm{GW}_{\mathrm{LW}} &\simeq O[(\omega L)^3]. \label{eq:mylwt1}
\end{align}


\begin{thebibliography}{99}
\bibitem{TDInote2} Hence, in contrast to Refs.\ \cite{STEA,TEA03}, we can ignore the difference between co-rotating and counter-rotating armlengths, and we can replace the ordered time delays ``;'' by the commuting time delays ``,''; finally, because we approximate the LISA array as a rotating equilateral triangle, we set $L_1 = L_2 = L_3 = L$.
\bibitem{notesub} The straightforward subtraction procedure used to remove the first source would be inadequate if more than a few sources of comparable strength were present in the signal: in that case, the interference caused by the additional sources would skew the accuracy of parameter estimation in a such a way that source-by-source subtraction would fail to converge. In the {\sc gCLEAN} procedure proposed in Ref.\ \cite{CL03}, subtraction proceeds iteratively by removing only a small fraction of the current brightest source before evaluating the parameters of the next brightest source.
\end{thebibliography}
\end{document}